\documentclass[journal]{IEEEtran}

\ifCLASSINFOpdf
\else
   \usepackage[dvips]{graphicx}
\fi
\usepackage{url}
\usepackage{hyperref}
\usepackage{cite}
\usepackage{amssymb}
\usepackage{pifont}
\newcommand{\cmark}{\ding{51}}%
\usepackage{mathtools}
\usepackage{subcaption} 
\usepackage{graphicx}
\usepackage{efbox}
\usepackage{multirow}
\usepackage{url}
\usepackage{xcolor}
\definecolor{newcolor}{rgb}{.8,.349,.1}

\usepackage{lineno,hyperref}

\begin{document}
\title{Q-matrix Unaware Double JPEG Detection \\using DCT-Domain Deep BiLSTM Network}

\author{Vinay Verma, Deepak Singh, and Nitin Khanna \\
	Multimedia Analysis and Security (MANAS) Lab, Electrical Engineering,\\
Indian Institute of Technology Gandhinagar (IITGN), Gujarat, India
\thanks{Affiliation: Electrical Engineering at Indian Institute of Technology Gandhinagar, Palaj, Gandhinagar, Gujarat, India, 382355 (e-mail:\{vinay.verma, nitin.khanna\}@iitgn.ac.in, deepak.singh@msc2016.iitgn.ac.in)}
}
\maketitle

\begin{abstract}

The double JPEG compression detection has received much attention in recent years due to its applicability as a forensic tool for the most widely used JPEG file format.
Existing state-of-the-art CNN-based methods either use histograms of all the frequencies or rely on heuristics to select histograms of specific low frequencies to classify single and double compressed images.
However, even amidst lower frequencies of double compressed images/patches, histograms of all the frequencies do not have distinguishable features to separate them from single compressed images.
This paper directly extracts the quantized DCT coefficients from the JPEG images without decompressing them in the pixel domain, obtains all AC frequencies' histograms, uses a module based on $1\times 1$ depth-wise convolutions to learn the inherent relation between each histogram and corresponding q-factor, and utilizes a tailor-made BiLSTM network for selectively encoding these feature vector sequences.
The proposed system outperforms several baseline methods on a relatively large and diverse publicly available dataset of single and double compressed patches.
Another essential aspect of any single vs. double JPEG compression detection system is handling the scenario where test patches are compressed with entirely different quantization matrices (Q-matrices) than those used while training; different camera manufacturers and image processing software generally utilize their customized quantization matrices.
A set of extensive experiments shows that the proposed system trained on a single dataset generalizes well on other datasets compressed with completely unseen quantization matrices and outperforms the state-of-the-art methods in both seen and unseen quantization matrices scenarios. 
\end{abstract}


\IEEEpeerreviewmaketitle

\section{Introduction}
\label{sec:djpeg_intro}

JPEG is a mainstream compression standard deployed in most image processing tools and inside camera processing modules.
Based on a user study in~\cite{bib:Park_2018_ECCV} that collected around 127k digital images over two years, found that majority of these images (77.95\%) were encoded in JPEG format.
Due to the broad usability of the JPEG compression standard, the forensics community, through the years, has given significant attention to JPEG compression-based forensics.
Double compression detection is an important forensic tool that can comment on JPEG image forensic history mainly in two ways. 
First, traces of double compression in a test image under consideration raise the possibility of the image being manipulated. 
While in another scenario, single vs. double compression detection tool helps in finding the tampered regions (if any) in a given image, where the single compressed blocks (parts of an image aligned with $8\times8$ grid) within a given JPEG image are indicative of the tampered regions~\cite{bib:lin2009fast,bib:Park_2018_ECCV}.
A comprehensive overview of other image forensics-related problems and the respective solutions can be found in~\cite{bib:piva2013overview,bib:stamm2013overview}.

In literature, double compression detection in JPEG images using blocking artifacts in the pixel domain and artifacts in the transform domain is explored. 
Luk{\'a}{\v{s}} \textit{et al.}~\cite{bib:lukas03} showed that for double compressed JPEG images, the histogram of discrete cosine transform (DCT) coefficients shows specific patterns such as local maximums, local minimums, and double peaks. 
In another seminal work, Popescu \textit{et al.}~\cite{bib:popescu04} demonstrated that the histogram of quantized DCT coefficients for the double compressed JPEG images exhibits periodic artifacts. 
In contrast, the corresponding histogram for single compressed JPEG images does not show such artifacts. 
Another subsequent study in~\cite{bib:lin2009fast} also demonstrated the periodic artifacts in the histogram of quantized DCT coefficients, and a probability model was proposed to localize original and forged regions.
Further, a better probability model for the DCT coefficients of a JPEG image was utilized in~\cite{bib:bianchi2011improved}, resulting in improved performance.  
The periodic artifacts of JPEG images in spatial as well as frequency domain were simultaneously modeled in~\cite{bib:chen2011tifs} for detecting recompression. 
In~\cite{bib:wang2014}, another approach for estimating the probability distribution of DCT coefficients was proposed for localizing the tampered region in an image, which showed improvement over~\cite{bib:lin2009fast} and~\cite{bib:bianchi2011improved}. 
Benford's law based approach has been used in~\cite{bib:fu07,bib:li08,bib:taimori17} for JPEG compression related forensics.
Other approaches based on handcrafting on the histogram of DCT coefficients were explored in~\cite{bib:feng2010jpeg,bib:pevny2008detection,bib:amerini14}.

Wang~\textit{et al.}~\cite{bib:wang2016double} was the first to use a 1D CNN (having 1D convolutions) with input as concatenated histogram obtained from the quantized DCT coefficients of the first nine AC frequencies in zig-zag order, resulting in better performance than handcrafted feature-based methods. 
Later, Barni~\textit{et al.}~\cite{bib:barni2017aligned} obtained further improvement in performance using three different strategies to train CNN. 
In one approach, the cumulative histogram of de-quantized DCT coefficients corresponding to all frequencies was estimated, the difference was computed and used as input to a CNN. 
In contrast, in the other two approaches, input to the CNN was image pixel values and denoised versions of the image. 
CNN trained with the histogram-based handcrafted features for aligned JPEG recompression gave a better performance than the other two approaches. 
A multi-branch network was used in~\cite{bib:li2017multi} to address the issue of double JPEG compression detection.
Authors in~\cite{bib:amerini2017localization} have used direct pixel domain image patches as input to a CNN, the histogram of DCT coefficients as input to another 1D CNN, and the combination of these two approaches.   
Another CNN based system in~\cite{bib:zeng2019detection} jointly trained a two branch parallel network; in one branch, pixel domain patches were passed to a specific filtering layer consisting of twelve fixed filters before passing the filtered volume to modified DenseNet. 
While in the other branch, which consists of two 1D convolutional layers and a fully connected layer, 1D DCT histograms obtained from the first nine frequencies in zig-zag order (excluding DC) were passed. 
Another approach in~\cite{bib:verma2018dct} selectively chose bin ranges for low frequencies from the luminance and the two chrominance channels and combined them to a 1D histogram before utilizing 1D CNN for multiple compression classification.
Authors in~\cite{bib:Park_2018_ECCV} derived features from the histogram of de-quantized DCT coefficients and used them as input to a CNN to learn to distinguish between single and double compressed blocks, and obtained better performances as compared to previous methods (~\cite{bib:wang2016double} and~\cite{bib:barni2017aligned}).
Further, this work~\cite{bib:Park_2018_ECCV} experimentally showed that concatenation of reshaped ($1\times64$) quantization matrices with the activation outputs of the last three dense layers resulted in improved performances.
One major deviation of this work from all other earlier work was the consideration of a more diverse set of 1120 quantization matrices (extracted from images collected using a forensic website) than 100 standard quantization matrices.
These compressed patches generated using these 1120 Q-matrices are made publicly available by the authors of~\cite{bib:Park_2018_ECCV}.
Another recent paper~\cite{bib:kwoncat} that focused on detecting image splicing utilized a double compression detection module as a part of their proposed splicing detection framework. 
First, a single vs. double compression classifier is separately trained on the same dataset by~\cite{bib:Park_2018_ECCV} and further used this pre-trained model in their splicing detection framework.
This single vs. double compression classification framework~\cite{bib:kwoncat} utilized an alternative preprocessing strategy in place of preprocessed histograms and designed input to the CNN using the concept of binary volume representation used in steganalysis~\cite{bib:yousfi2020intriguing}. 
Although only using the binary volume representation of DCT coefficients gave slightly lower accuracy than the existing method (91.71\%~\cite{bib:kwoncat} vs 92.76\%~\cite{bib:Park_2018_ECCV}, Table 1 in~\cite{bib:kwoncat}).
However, when this binary volume representation of DCT coefficients is combined with utilization of q-factors at the respective frequency locations in binary volume processed feature map, the combined system outperformed existing state-of-art (93.93\%~\cite{bib:kwoncat} vs. 92.76\%~\cite{bib:Park_2018_ECCV}, Table 1 in~\cite{bib:kwoncat}). 
Moreover, the idea of using q-factors with the respective histograms at the input design side is also presented in our earlier work~\cite{bib:verma2020block}. 
In this paper, we have further extended the idea of using q-factors with the respective histograms proposed in~\cite{bib:verma2020block} and shown that our proposed approach outperforms existing state-of-the-art methods.

One primary observation with the histogram-based CNN methods is that, in all the CNN based method using histograms as input, either coefficients (quantized/de-quantized) from few selected low frequency indexes (in zig-zag order)~\cite{bib:wang2016double,bib:amerini2017localization,bib:verma2018dct} or the coefficients from all the frequency locations~\cite{bib:barni2017aligned,bib:Park_2018_ECCV} are utilized.
However, heuristically choosing specific frequency locations or using coefficients from all the frequency locations might not be an optimal approach for histogram-based CNN methods for double JPEG compression detection due to the following reasons.
Selecting only specific frequency locations for the histogram formation is susceptible to miss information in the coefficients from the high-frequency locations.
While in another scenario, where coefficients from all frequency locations are utilized for histogram formation, it has the following drawback. 
For all the frequency locations, even amidst the lower frequencies, not all pairs of first and second q-factors result in distinguishable histograms (Section~\ref{subsec:inp_design}).
In this paper, we tailored a long short-term memory (LSTM) based system for single vs. double compression detection that selectively encodes histogram-based features obtained from all the AC frequencies.
First, histograms are formulated for each AC frequency. 
These histograms and respective q-factors are passed to a designed module termed as a non-linear HQ projector ($\phi(\cdot)$) that captures; the histograms' inherent relationship with the corresponding q-factors.
Further, the output feature sequence obtained from module $\phi(\cdot)$, corresponding to 63 AC frequencies, is fed to a deep BiLSTM, tailored explicitly for the task, to selectively encode all 63 histogram feature sequences into a single fixed-size vector before the final classification.
For the histogram formation, we used quantized DCT coefficients directly extracted from JPEG bit-stream.

For the performance evaluation, we consider a diverse public dataset also considered in the recent studies~\cite{bib:Park_2018_ECCV,bib:kwoncat}. 
This dataset is derived from several uncompressed datasets captured with several cameras (Section~\ref{sec:djpeg_exp}) and has nearly 1.14 million $256\times 256$ single and double compressed patches, using 1,120 quantization matrices.
This complex and diverse dataset provides two advantages. 
First, it enables the design of a single classifier to distinguish between single and compressed JPEG images compared to creating multiple classifiers based on the last quantization matrix of the test image as done traditionally~\cite{bib:pevny2008detection,bib:chen2011tifs,bib:barni2017aligned,bib:amerini2017localization,bib:verma2018dct}.  
Second, this dataset can be used for standard benchmarking of the current and future methods.  
Furthermore, 1,120 quantization matrices are more extensive than the 100 standard ones; they still are not a closed set; different camera manufacturers and image processing software generally use their customized quantization tables.
Any single vs. double compression method should handle the scenario where the test patch is compressed with an entirely different Q-matrix compared to what was used for training.
To test the Q-matrix generalizability, we explore four standard uncompressed datasets and evaluate them in seen/unseen Q-matrices scenarios.

Experimental results demonstrate the proposed approach's effectiveness with robust performances in image content and quantization-matrix unaware scenarios.
To summarize, the following are the major contributions of this paper:

\begin{itemize}

\item We attempt to solve the problem of single vs. double compression detection from a new perspective. 
In this direction, for the first time, we design a bidirectional LSTM based system that allows effective encoding of the features obtained from the histograms of quantized DCT coefficients and respective quantization factors and outperforms current state-of-the-art methods.

\item We designed a $1\times 1$ convolution based non-linear HQ projector module that combines raw quantized DCT histograms and respective quantization factors in a novel way to provide input for the proposed BiLSTM network.

\item Significant performance improvement over the next best method for classifying smaller sized patches (5.74\% for $64\times64$, 5.24\% for $128\times128$ in test accuracy) (Section~\ref{subsec:patch_perf}).

\item Proposed BiLSTM based framework outperforms state-of-the-art convolutional neural network-based methods on standard datasets, especially for unseen datasets and unseen quantization matrices.  
It depicts the proposed method's generalization capability and its applicability for unseen patches compressed with unseen quantization matrices. 
\end{itemize}

\section{Proposed System}
\label{sec:djpeg_prop_sys}

\begin{figure*}[!htb]
    \centering
    \includegraphics[width = \textwidth,trim={0 0 10cm 0},clip]{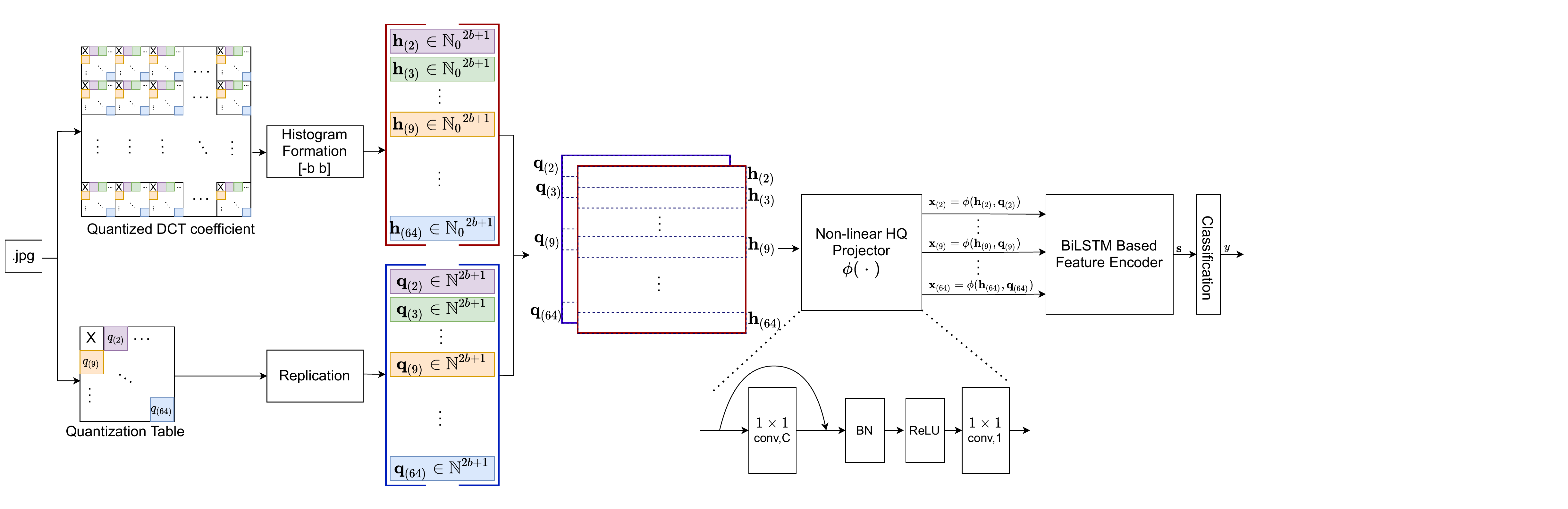}
    \caption{Proposed framework. It consists of three main parts: input design, a non-linear HQ projector module $\phi(.)$, and a BiLSTM based feature encoder. 
$b$ is the bin range parameter for histogram formation. $\mathbf{h}_k$ and $\mathbf{q}_k$ are the histogram vector and replicated q-factor vector at frequency location $k$. Different colors here refer to different frequency locations, $k\in\{2,3,\ldots,64\}$, while `x' represents that DC frequency is excluded.  
In module $\phi(.)$, $1\times1$, $C$ refers to $1\times1$ convolution with $C$ filters, BN refers to batch normalization, and ReLU refers to rectified linear unit. 
$\mathbf{x}_2,\;\mathbf{x}_3,\;\ldots,\;\mathbf{x}_{64}$ is the sequence of feature vectors passed to BiLSTM based encoder to obtain a final feature vector for classification.}
    \label{fig:block_dia}
\end{figure*}

Figure~\ref{fig:block_dia} shows the overall framework of the proposed system.
This framework consists of three major components: an input design part, a non-linear HQ projector module $\phi(.)$, and a feature encoder. 
We first present a brief background of JPEG compression, and subsequent sections describe details of these three components.

\begin{figure}[!htb]
    \centering
    \includegraphics[width = 0.5\textwidth,trim={1.3cm 0 1cm 0},clip]{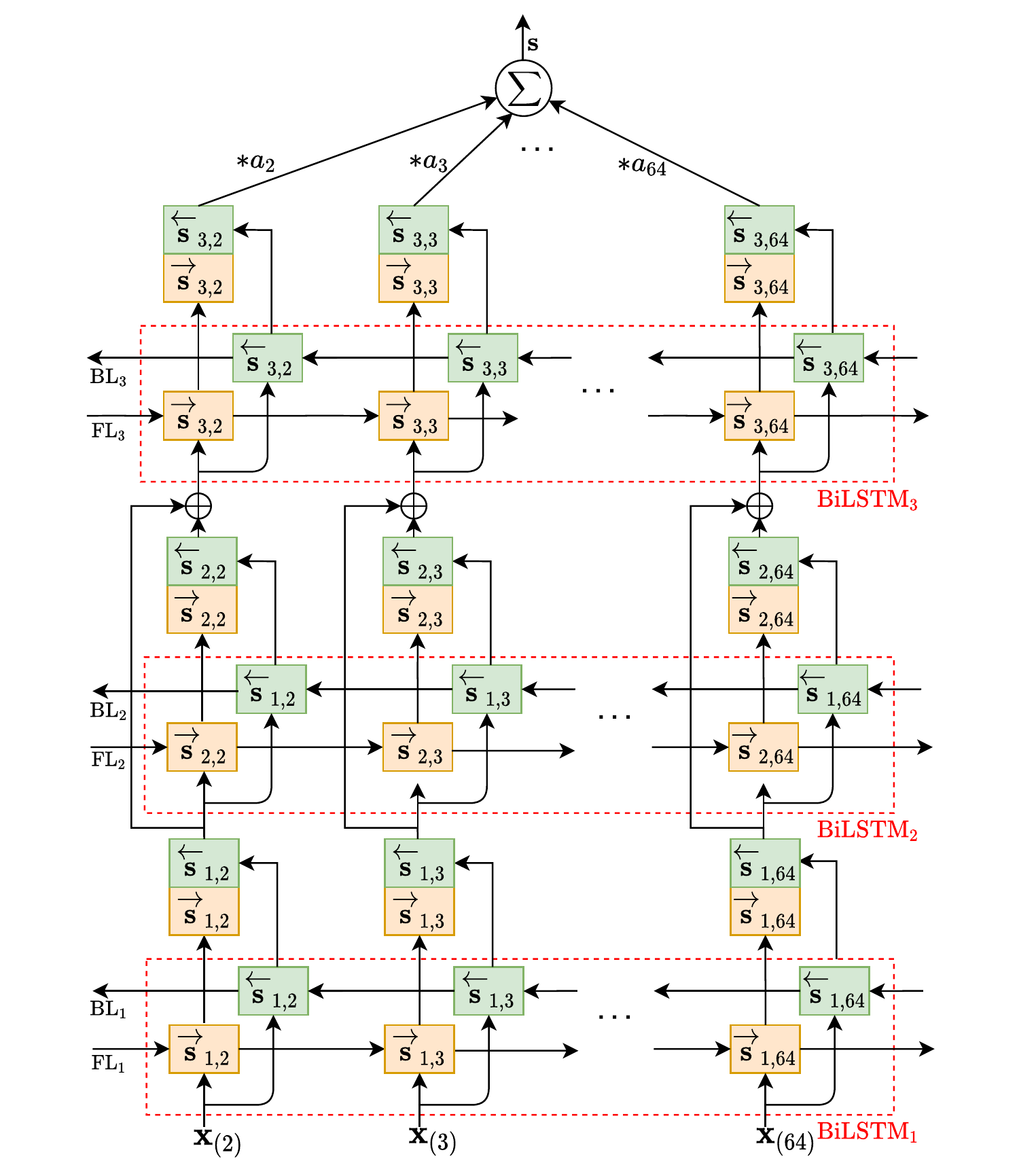}
    \caption{BiLSTM based feature encoder: It consists of three BiLSTM blocks (BiLSTM$_1$, BiLSTM$_2$, and BiLSTM$_3$) each consisting of one forward layer (FL) and one backward layer (BL). $\protect\overrightarrow{\mathbf{s}}_{l,k}$ and $\protect\overleftarrow{\mathbf{s}}_{l,k}$ represent forward hidden state and backward hidden state vectors at layer $l\in\{1,2,3\}$ and frequency index $k\in\{2,3,\ldots,64\}$. These output hidden states are concatenated $([\protect\overrightarrow{\mathbf{s}}_{l,k};\protect\overleftarrow{\mathbf{s}}_{l,k}])$ at each frequency index $k$ and layer $l$ before further processing. $a_k$'s ($k\in\{2,3,\ldots,64\}$) are the scalar weights.}
    \label{fig:network}
\end{figure}

An RGB image's JPEG compression~\cite{bib:wallace1992jpeg} begins with a change of color space, conversion into luminance-chrominance channels (YCbCr). 
Each of these three channels is partitioned into non-overlapping blocks of size $8\times8$, and 2D discrete cosine transform (DCT) coefficients are estimated for each of these blocks and independently quantized.
For the quantization purpose, the transformed blocks are quantized with an $8\times8$ quantization matrix (Q-matrix).
A separate Q-matrix can be chosen for each of the three channels; however, all $8\times 8$ transformed blocks of a specific channel are quantized with the same Q-matrix.
Entropy encoding of the quantized coefficients results in JPEG encoded bit-stream. 
JPEG decompression involves applying the compression steps in reverse order; entropy decoding, dequantization, inverse DCT (IDCT).
Rounding and truncation operation is applied on the obtained IDCT values to get the image pixels in the nearest integers.

\subsection{Input Design}
\label{subsec:inp_design}

The proposed work builds on the fact that for aligned double JPEG compression with varying quantization matrices, histograms of quantized DCT coefficients have different statistical properties, based on the number of times an image has been JPEG compressed~\cite{bib:lukas03, bib:popescu04, bib:lin2009fast, bib:chen2011tifs, bib:barni2017aligned}. 
Let $\{U(k_1,k_2)\}$ denotes the collection of DCT coefficients of an uncompressed image, at a frequency location $(k_1,k_2)$ ($k_1,k_2\in \left\{1,2,\ldots8\right\}$).
For a single compressed image, the collection of quantized DCT coefficients is $\left\{\left[\dfrac{U(k_1,k_2)}{Q_1(k_1,k_2)}\right]\right\}$, where $Q_1$ is ${8\times 8}$ quantization matrix (Q-matrix) used for single compression, and $Q_1(k_1,k_2)$ is the quantization step size (q-factor) at frequency $(k_1,k_2)$ ($[.]$ denotes the rounding operation). 
For a double compressed image, DCT coefficients of an uncompressed image are quantized during the first compression, de-quantized with $Q_1(k_1,k_2)$, and further quantized with $Q_2(k_1,k_2)$ during the second JPEG compression.
So, for double compressed image, the collection of quantized DCT coefficients at $(k_1,k_2)$ is $\left\{\left[\left[\dfrac{U(k_1,k_2)}{Q_1(k_1,k_2)}\right]\dfrac{Q_1(k_1,k_2)}{Q_2(k_1,k_2)}\right]\right\}$.
Note that in this formulation, the effect of pixel rounding and truncation during the JPEG decompression process is neglected.

Let the unquantized DCT coefficients (DCT coefficients of an uncompressed natural image), at a particular AC frequency $(k_1,k_2)$ ($(k_1,k_2) \neq (1, 1)$) are denoted as a random variable $U$. 
In literature, the distribution of $U$ has been generally modeled as Gaussian or Laplacian distribution~\cite{bib:lam2000mathematical, bib:wang2014}.
For our demonstration purposes, let the generalized probability density function of $U$ be represented as $f_{U}(u;\theta)$, $\theta$ being the distribution's parameter. 
The system design proposed in this paper does not depend upon the exact form of $f_U$ and is equally valid for a general probability density function $f_U$. We have also validated this by initial experiments on 1 million randomly drawn numbers 
from uniform, Gaussian, and Laplacian distributions. 
Further, let $S$ denotes the random variable corresponding to the quantized DCT coefficients corresponding to single compressed images quantized by a q-factor $q_1$ ($q_1$ will correspond to specific $Q_1(k_1,k_2)$, for a particular value of $(k_1,k_2)$). Then, operations related to JPEG compression can be modeled in terms of functions of random variables. 
\begin{equation*}
    S = \left[{\dfrac{U}{q_1}}\right]\implies q_1\left(S - 0.5\right)\leq U <q_1\left(S + 0.5\right) 
\end{equation*}

Thus, the distribution (probability mass function) of $S$ can be written as: 
\begin{equation}
    P_{S}(s) = \int_{ q_1\left(s - 0.5\right)}^{ q_1\left(s + 0.5\right)}f_{U}(u;\theta) du
    \label{eq:dist_single}
\end{equation}

Consider another random variable $D$ denoting the quantized DCT coefficients corresponding to the double compressed JPEG images, that is, unquantized DCT coefficients first quantized by $q_1$ then by $q_2$. 
\begin{equation*}
    D = \left[{\left[{\dfrac{U}{q_1}}\right]\dfrac{q_1}{q_2}}\right] \implies
\end{equation*}
\begin{equation*}
\resizebox{0.91\hsize}{!}{%
        $q_1\left(\left\lceil{\dfrac{q_2}{q_1}\left(D - 0.5\right)}\right\rceil -0.5\right) \leq U < q_1\left(\left\lceil{\dfrac{q_2}{q_1}\left(D + 0.5\right)}\right\rceil -0.5\right)$%
        }
\end{equation*}

The probability mass function of $D$ can be written as: 
\begin{equation}
    P_{D}(d) = \int_{q_1\left(\left\lceil{\dfrac{q_2}{q_1}\left(d - 0.5\right)}\right\rceil -0.5\right)}^{ q_1\left(\left\lceil{\dfrac{q_2}{q_1}\left(d + 0.5\right)}\right\rceil -0.5\right)}f_{U}(u;\theta)du
    \label{eq:dist_double}
\end{equation}

Depending upon the relative values of $q_1$ and $q_2$ ($q_1,q_2\in\{1,2,\ldots,255\}$~\cite{bib:wallace1992jpeg}), certain combinations of $q_1$ and $q_2$ introduce specific artifacts in the distribution of double quantized DCT coefficients. 
These artifacts help in single and double compression detection.
For a double compressed image, at each frequency location $(k_1,k_2)$, one of the following five scenarios can occur~\cite{bib:lukas03,bib:farid2016photo}:  

\begin{enumerate}
    \item $q_1>q_2$, and $q_1$ is an integer multiple of $q_2$
    \item $q_1>q_2$, and $q_1$ is not an integer multiple of $q_2$
    \item $q_1<q_2$, and $q_2$ is an integer multiple of $q_1$
    \item $q_1<q_2$, and $q_2$ is not an integer multiple of $q_1$
    \item $q_1=q_2$
\end{enumerate}

For both scenarios 1 and 2, the distribution of double quantized DCT coefficients ($P_D$) has missing bins (for several values of d, $P_D(d) = 0$). 
For scenario 4, the distribution of double quantized DCT coefficients has extremas (peak-valley patterns). 
These peculiarities in the distribution of double quantized DCT coefficients help in distinguishing them from single quantized DCT coefficients, which do not have such artifacts in the distribution of single compressed coefficients. 
However, for scenarios 3 and 5, the distributions of double quantized DCT coefficients do not contain observable compression artifacts that can separate them from single compression. 

Now, based on the above discussion, and Equation~\ref{eq:dist_single} and Equation~\ref{eq:dist_double}, we can conclude the following:
\begin{enumerate}
    \item  Distribution of the single quantized DCT coefficients depends on $q_1$ and on the distribution of the original unquantized DCT coefficients ($f_{U}(u;\theta)$) .
    \item Distribution of the double quantized DCT coefficients depends on $q_1$, $q_2$ and on the distribution of the original unquantized DCT coefficients ($f_{U}(u;\theta)$).
    \item Out of 64 distributions of double quantized DCT coefficients, not all distributions have distinguishable statistical characteristics based on the relative values of $q_1$ and $q_2$ at different frequency locations.
\end{enumerate}

To build on this fact, we propose using q-factors available in the JPEG header with the corresponding raw histograms of the quantized DCT coefficients for each of the AC frequencies at the input side of the proposed network.
The majority of the double compression based methods such as~\cite{bib:li08,bib:pevny2008detection,bib:chen2011tifs,bib:wang2014,bib:wang2016double,bib:amerini2017localization,bib:zeng2019detection} neglect the coefficients from DC frequency ($k_1=1, k_2=1$).
For a JPEG image, its header always contains Q-matrix used during the final compression ($Q_1$ for the single compressed image, and $Q_2$ for the double compressed image). 
Although for a double compressed patch, we do not have access to the first Q-matrix $Q_1$, the relationship of q-factors with the corresponding quantized DCT coefficients is captured in the corresponding histogram~\cite{bib:lin2009fast}.
We design a module termed as a non-linear projector ($\phi(\cdot)$) that uses a combination of depth-wise convolutions (details in Section~\ref{subsec:hist_q_reln}) with the histograms and corresponding q-factors to learn the inherent relationships between them. 
Further, the output of module ($\phi(\cdot)$) is utilized as a sequence corresponding to 63 AC frequencies, which is used as input to our BiLSTM based feature encoder (Section~\ref{subsec:bi_feat_encoder}) to selective encode the feature sequence into a fixed length vector for the classification task.

In general a frequency location for each $8\times8$ DCT coefficients is denoted by the tuple of horizontal and vertical frequencies $(k_1,k_2)$ ($k_1,k_2\in \left\{1,2,\ldots8\right\}$).
Here for compactness of notations further used in the paper, we first convert from 2d notation $(k_1,k_2)$ to the corresponding scalar notation $k$ ($k=1,2,3,\ldots,64$) in raster order, defined in Equation~\ref{eq:frq_2d_1d}.
\begin{equation}
k = (k_1-1)*8 + k_2, 
\label{eq:frq_2d_1d}
\end{equation} 
Now for going back to the 2d frequency $(k_1,k_2)$, the following set of rules described in Equation~\ref{eq:frq_1d_2d} (where $(k)_8$ denotes k modulo 8 or remainder obtained after dividing k by 8) can be used.
\begin{equation}
(k_1,k_2) = \begin{cases}
  \left(\dfrac{k}{8}, 8\right), & \text{if } (k)_8= 0, \\
  \left(\left\lfloor{\dfrac{k}{8}}\right\rfloor+1, (k)_8\right), & \text{otherwise.}
\end{cases}
\label{eq:frq_1d_2d}
\end{equation}

Quantized DCT coefficients are obtained directly from the JPEG file's bit-stream without fully decompressing the JPEG image.
Irrespective of the patch being single or double compressed, let $\left\{F_{q}(k)\right\}$ denote its collection of quantized DCT coefficients at a frequency location $k$.
Histogram of quantized DCT coefficients at frequency location $(k)$ with integer bins in the range $[-b\;b]$, is defined as: $h_{(k)}(i) = |\{F_q(k) | F_q(k) = i\}|$, $i \in \left[-b\;b \right]$ ($|.|$ is cardinality of a set). 
Hence, histogram for a particular frequency $\mathbf{h}_{(k)}\in \mathbb{N}_0^{(2b+1)}$, where $\mathbb{N}_0$ (set of all non-negative integers) $=\mathbb{N}\cup\{0\}$.
Further, the utilization of the available q-factor for each of the frequency $k$ with their respective histograms is done in a specific way, as described in the following section.

\subsection{Non-linear HQ Projector
($\phi(\mathbf{h}_{(k)}, \mathbf{q}_{(k)})$)} 
\label{subsec:hist_q_reln}

For jointly utilizing the information present in the histogram and the q-factor, we have proposed a non-linear projector module that involves the following set of operations.
First, $C$, $1\times1$ convolutions with linear activation are used to get $C$ linear projected planes. 
The linear projected value for a particular filter $c$, at frequency index $k$, and histogram bin index $i$, is defined as $h_{(k,i,c)}^{P}$ (Equation~\ref{eq:c_lin_proj}).
\begin{equation}
\label{eq:c_lin_proj}
\begin{split}
 h_{(k,i,c)}^{P} = \alpha_{c}h_{(k,i)} + \beta_{c}q_{k} +\gamma_c, \; \forall k\in\{2,3,\ldots,64\},
 \\\forall i\in\{-b,-b+1,\ldots,b\},\;\forall c\in\{1,2,\ldots,C\}.
\end{split}
\end{equation}
Here $h_{(k,i)}$ is number of quantized DCT coefficients for the frequency location $k$ at histogram bin index $i$, while $q_{k}$ is q-factor at frequency index $k$ ($Q(k)$). $\alpha_{c}$ and  $\beta_{c}$ are trainable weights, while $\gamma_c$ is the bias value corresponding to $c^{th}$ filter ($c\in\{1,2,\ldots,C\}$).   
To achieve projections as defined in Equation~\ref{eq:c_lin_proj} using $1\times1$ convolutions, q-factor values at each frequency index $k$ are repeated for the entire histogram bin range so that $q_k=q_{(k,i)},\; \forall i\in\{-b,-b+1,\ldots,b\}$.
This implies, for each frequency $(k)$, corresponding q-factor $q_k$ is repeated $(2b+1)$ times to obtain $\mathbf{q}_{(k)}\in \mathbb{N}^{(2b+1)}$ and channel-wise concatenated with the respective histograms $\mathbf{h}_{(k)}\in \mathbb{N}_0^{(2b+1)}$ to get the representation of dimension $1\times(2b+1)\times2$ at each AC frequency.
Further, these $C$ linear projected channels (planes) are concatenated with the input resulting in $C+2$ channels before passing to a batch normalization layer, ReLU non-linearity, and a $1\times 1$ convolution layer to get the feature representations.
The complete order of operations is represented as a non-linear HQ projector ($\phi(.)$) in Figure~\ref{fig:block_dia}.
Obtained feature representations corresponding to 63 AC frequencies are used as sequence $\mathbf{x}_2,\;\mathbf{x}_3,\;\ldots,\mathbf{x}_{64}\in \mathbb{R}^{2b+1}$ for further processing with BiLSTM based feature encoder described in the following section.

\subsection{BiLSTM-based Feature Encoder}
\label{subsec:bi_feat_encoder}

We design a long short-term memory (LSTM)~\cite{bib:hochreiter1997long} based model to selectively encode the obtained obtained feature sequences effectively for the classification of single and double compressed JPEG patches. 
LSTM based methods have been successfully used to improve state-of-the-art for tasks related to sequential data such as language modeling, time series prediction, sentiment analysis, text recognition, and many other applications~\cite{bib:van2020review}. 
LSTM is a type of recurrent neural network that addresses the vanishing gradient problem in the vanilla RNN~\cite{bib:hochreiter1997long}.
The core idea behind the LSTM cells is a cell state that runs through the entire sequence indices and gate units that regulate the information flow into and out of the cell state at every step of processing sequential input.
At every step of the frequency index, each LSTM cell maintains two states, a cell state $\overrightarrow{\mathbf{c}}_{k}$ and a hidden state $\overrightarrow{\mathbf{s}}_{k}$. 
Here `$\overrightarrow{}$' denotes forward processing of the sequence.
The cell state is internal to the LSTM cell, and the hidden state is revealed as the LSTM cell's output. 
Let $\overrightarrow{\mathbf{c}}_{k-1}$ and $\overrightarrow{\mathbf{s}}_{k-1}$ be the previous cell state and hidden state, respectively. 
The current cell-state $\overrightarrow{\mathbf{c}}_{k}$ and current hidden state $\overrightarrow{\mathbf{s}}_{k}$ is regulated by gated units based on the current input $\mathbf{x}_{k}$, previous cell state $\overrightarrow{\mathbf{c}}_{k-1}$, and previous hidden state $\overrightarrow{\mathbf{s}}_{k-1}$.

As mentioned above, an LSTM cell consists of several control gates such as a forget gates, two input gates, and an output gate that regulate the cell state and provide current hidden state.
Forget gate regulates what information need to be forgetted or remembered from the previous cell state $\overrightarrow{\mathbf{c}}_{k-1}$, using the element-wise multiplication of forget gate activation vector $\overrightarrow{\mathbf{f}}_k$ (Equation~\ref{eq:fk}) with $\overrightarrow{\mathbf{c}}_{k-1}$ ($\overrightarrow{\mathbf{f}}_k \odot \overrightarrow{\mathbf{c}}_{k-1}$).
Further, another set of input gates regulates what new information needs to be added in the cell state.
This is a two step process, first the input gate activation vector $\overrightarrow{\mathbf{i}}_k$ (Equation~\ref{eq:ik}) decides which values to update and then another input gate activation, named as candidate activation vector $\overrightarrow{\widetilde{\mathbf{c}}}_k$ (Equation~\ref{eq:c_til_k}) is generated, and finally these two vectors are element-wise multiplied ($\overrightarrow{\mathbf{i}}_k \odot \overrightarrow{\widetilde{\mathbf{c}}}_k$). 
Thus, the current cell state $\overrightarrow{\mathbf{c}}_{k}$ is updated as described in Equation~\ref{eq:ck}.
An output gate activation vector $\overrightarrow{\mathbf{o}}_k$ (Equation~\ref{eq:ok}) regulates what part of the information to pass from the $\tanh$ squashed version of the current cell state ($\tanh(\overrightarrow{\mathbf{c}}_{k})$) to current hidden state $\overrightarrow{\mathbf{s}}_k$. 
Thus, the current hidden state $\overrightarrow{\mathbf{s}}_k$ is updated using the element-wise multiplication of output gate activation vector $\overrightarrow{\mathbf{o}}_k$ with the cell state values obtained after passing through $\tanh$ activation, as described in Equation~\ref{eq:sk}. 
The $\sigma(.)$, $\tanh(.)$, and `$\odot$' denote the sigmoid, hyperbolic tangent, and element wise multiplication functions, respectively. 
\begin{align}
    \overrightarrow{\mathbf{f}}_k & = \sigma(\overrightarrow{\mathbf{W}}_{f}\mathbf{x}_{k} + \overrightarrow{\mathbf{V}}_f\overrightarrow{\mathbf{s}}_{k-1} + \overrightarrow{\mathbf{b}}_f)
    \label{eq:fk}\\
    \overrightarrow{\mathbf{i}}_k & = \sigma(\overrightarrow{\mathbf{W}}_i\mathbf{x}_{k} + \overrightarrow{\mathbf{V}}_i\overrightarrow{\mathbf{s}}_{k-1} + \overrightarrow{\mathbf{b}}_i)
    \label{eq:ik}\\
    \overrightarrow{\mathbf{o}}_k & = \sigma(\overrightarrow{\mathbf{W}}_o\mathbf{x}_{k} + \overrightarrow{\mathbf{V}}_o\overrightarrow{\mathbf{s}}_{k-1} + \overrightarrow{\mathbf{b}}_o) \label{eq:ok}\\
    \overrightarrow{\widetilde{\mathbf{c}}}_k & = \tanh(\overrightarrow{\mathbf{W}}_c\mathbf{x}_{k} + \overrightarrow{\mathbf{V}}_c\overrightarrow{\mathbf{s}}_{k-1} + \overrightarrow{\mathbf{b}}_c) \label{eq:c_til_k}\\
    \overrightarrow{\mathbf{c}}_k & = \overrightarrow{\mathbf{f}}_k \odot \overrightarrow{\mathbf{c}}_{k-1} + \overrightarrow{\mathbf{i}}_k \odot \overrightarrow{\widetilde{\mathbf{c}}}_k \label{eq:ck}\\ 
    \overrightarrow{\mathbf{s}}_k & = \overrightarrow{\mathbf{o}}_k \odot \tanh(\overrightarrow{\mathbf{c}}_k)\label{eq:sk}
\end{align}

Here, $\overrightarrow{\mathbf{W}}_{f},\overrightarrow{\mathbf{W}}_{i},\overrightarrow{\mathbf{W}}_{o},\overrightarrow{\mathbf{W}}_{c}\in \mathbb{R}^{n\times m}$ ($n$: number of hidden units of each LSTM cell, $m$: dimensionality of the input vector $\mathbf{x}_{k}$) are the input weight matrices related to the current input vector $\mathbf{x}_{k}$ and the respective gates.
$\overrightarrow{\mathbf{V}}_{f},\overrightarrow{\mathbf{V}}_{i},\overrightarrow{\mathbf{V}}_{o},\overrightarrow{\mathbf{V}}_{c}\in \mathbb{R}^{n\times n}$ are the recurrent weight matrices related to the previous hidden state vector $\overrightarrow{\mathbf{s}}_{k-1}$ and the respective gates.
$\overrightarrow{\mathbf{b}}_{f},\overrightarrow{\mathbf{b}}_{i},\overrightarrow{\mathbf{b}}_{o},\overrightarrow{\mathbf{b}}_{c}\in \mathbb{R}^{n}$ are the bias vectors related to the respective gates. 
For compactness purposes, we model these set of operations to obtain current hidden state vector $\overrightarrow{\mathbf{s}}_k $ from the previous hidden state vector $\overrightarrow{\mathbf{s}}_{k-1}$, current input $\mathbf{x}_{k}$, and set of gate parameters $\overrightarrow{\mathbf{\Theta}}:(\overrightarrow{\mathbf{W}}_{f},\overrightarrow{\mathbf{V}}_f, \overrightarrow{\mathbf{b}}_f,\overrightarrow{\mathbf{W}}_{i},\overrightarrow{\mathbf{V}}_i, \overrightarrow{\mathbf{b}}_i,\overrightarrow{\mathbf{W}}_{o},\overrightarrow{\mathbf{V}}_o, \overrightarrow{\mathbf{b}}_o,\overrightarrow{\mathbf{W}}_{c},\overrightarrow{\mathbf{V}}_c, \overrightarrow{\mathbf{b}}_c)$ with a function $g(.)$, denoted as $\overrightarrow{\mathbf{s}}_k = g(\overrightarrow{\mathbf{s}}_{k-1},\mathbf{x}_{k};\overrightarrow{\mathbf{\Theta}})$.

The above description of an LSTM cell corresponds to the forward LSTM. Similar computation $g(.)$ is performed in the backward LSTM with its own gate parameters $\overleftarrow{\mathbf{\Theta}}:(\overleftarrow{\mathbf{W}}_{f},\overleftarrow{\mathbf{V}}_f, \overleftarrow{\mathbf{b}}_f,\overleftarrow{\mathbf{W}}_{i},\overleftarrow{\mathbf{V}}_i, \overleftarrow{\mathbf{b}}_i,\overleftarrow{\mathbf{W}}_{o},\overleftarrow{\mathbf{V}}_o, \overleftarrow{\mathbf{b}}_o,\overleftarrow{\mathbf{W}}_{c},\overleftarrow{\mathbf{V}}_c, \overleftarrow{\mathbf{b}}_c)$. 
Here `$\overleftarrow{}$' denotes values related to the backward LSTM layer.
Let $\overleftarrow{\mathbf{s}}_k$ denotes the backward hidden state at frequency index $k$. Then based on the previous hidden state vector $\overleftarrow{\mathbf{s}}_{k+1}$ and current input $\mathbf{x}_{k}$ in the backward layer, we get $\overleftarrow{\mathbf{s}}_k = g(\overleftarrow{\mathbf{s}}_{k+1}, \mathbf{x}_{k}; \overleftarrow{\mathbf{\Theta}})$. 
Note that the forward layer parameter $\overrightarrow{\mathbf{\Theta}}$ and the backward layer parameter $\overleftarrow{\mathbf{\Theta}}$ are shared over the frequency indices.

Based on the above description of an LSTM cell, due to its capability to selectively decide what new information to incorporate or forget previous information at each frequency indices, LSTM cells are an appropriate choice for encoding the feature sequence.
Figure~\ref{fig:block_dia} shows the overall flow in the proposed system, while Figure~\ref{fig:network} further details the BiLSTM based feature encoder module. 
Bidirectional LSTM (BiLSTM)~\cite{bib:schuster1997bidirectional} has been used as a basic building block of the proposed network.
BiLSTM consists of two separate hidden layers named a forward layer LSTM and a backward layer LSTM.
At any frequency state $k$, the forward layer computes the hidden state $\overrightarrow{\mathbf{s}}_k$ by iterating over the frequency index 2 to 64. In contrast, the backward layer computes the hidden state $\overleftarrow{\mathbf{s}}_k$ by iterating over the frequency index 64 to 2 in backward order. 
Finally, at that frequency index $k$, both forward and backward hidden states are concatenated to get the representation at the index $k$.

Further, we designed a three-layer deep bidirectional LSTM architecture with a residual connection between the output of first layer and second layer at each frequency index.
At BiLSTM$_1$, the input sequence is $\mathbf{x}_k\in \mathbb{R}^{2b+1}$ ($k=2,3,\ldots,64$). Forward layer 1 (FL$_1$), provides the hidden state $\overrightarrow{\mathbf{s}}_{1,k}\in \mathbb{R}^n$, while backward layer 1 (BL$_1$), provides the hidden state $\overleftarrow{\mathbf{s}}_{1,k}\in \mathbb{R}^n$:
\begin{align*}
    \overrightarrow{\mathbf{s}}_{1,k} = g(\overrightarrow{\mathbf{s}}_{1,k-1}, \mathbf{x}_{k}, \overrightarrow{\mathbf{\Theta_1}} )\\
    \overleftarrow{\mathbf{s}}_{1,k} = g(\overleftarrow{\mathbf{s}}_{1,k+1}, \mathbf{x}_{k}, \overleftarrow{\mathbf{\Theta_1}} )
\end{align*}
Here $\overrightarrow{\mathbf{\Theta_1}}$ and $\overleftarrow{\mathbf{\Theta_1}}$ are the parameters for the FL$_1$ and BL$_1$, respectively.
Now at each frequency index $k$, forward and backward LSTM cell's hidden state are concatenated, and the resulting vector is denoted as $\mathbf{s}_{1,k}$ ($\mathbf{s}_{1,k} = [\overrightarrow{\mathbf{s}}_{1,k};\overleftarrow{\mathbf{s}}_{1,k}] \in \mathbb{R}^{2n}$).

At BiLSTM$_2$, the input sequence is $\mathbf{s}_{1,k}\in \mathbb{R}^{2n}$ ($k=2,3,\ldots,64$), and the corresponding forward ($\overrightarrow{\mathbf{s}}_{2,k}$) and backward ($\overleftarrow{\mathbf{s}}_{2,k}$) hidden state vector are represented below:
\begin{align*}
    \overrightarrow{\mathbf{s}}_{2,k} = g(\overrightarrow{\mathbf{s}}_{2,k-1}, \mathbf{s}_{1,k}; \overrightarrow{\mathbf{\Theta_2}} )\\
    \overleftarrow{\mathbf{s}}_{2,k} = g(\overleftarrow{\mathbf{s}}_{2,k+1}, \mathbf{s}_{1,k}; \overleftarrow{\mathbf{\Theta_2}} )
\end{align*}
Here $\overrightarrow{\mathbf{\Theta_2}}$ and $\overleftarrow{\mathbf{\Theta_2}}$ are the parameters for the FL$_2$ and BL$_2$, respectively.
Further, for each frequency index, the concatenated output hidden states of BiLSTM$_2$ $\mathbf{s}_{2,k}  = [\overrightarrow{\mathbf{s}}_{2,k};\overleftarrow{\mathbf{s}}_{2,k}] \in \mathbb{R}^{2n}$ is element-wise added with the $\mathbf{s}_{1,k}$ to obtain $\mathbf{s'}_{2,k}$ ($\mathbf{s'}_{2,k}  = [\mathbf{s}_{1,k} \oplus \mathbf{s}_{2,k}]$).

At BiLSTM$_3$, the input sequence is $ \mathbf{s'}_{2,k}\in \mathbb{R}^{2n}$, and the corresponding output:
\begin{align*}
    \overrightarrow{\mathbf{s}}_{3,k} = g(\overrightarrow{\mathbf{s}}_{3,k-1}, \mathbf{s'}_{2,k}; \overrightarrow{\mathbf{\Theta_3}} )\\
    \overleftarrow{\mathbf{s}}_{3,k} = g(\overleftarrow{\mathbf{s}}_{3,k+1}, \mathbf{s'}_{2,k}; \overleftarrow{\mathbf{\Theta_3}} )
\end{align*}

Here $\overrightarrow{\mathbf{\Theta_3}}$ and $\overleftarrow{\mathbf{\Theta_3}}$ are the parameters for the FL$_3$ and BL$_3$, respectively. Finally, $\mathbf{s}_{k}$ is obtained with the concatenation of $\overrightarrow{\mathbf{s}}_{3,k}$ and $\overleftarrow{\mathbf{s}}_{3,k}$ ($   \mathbf{s}_{k}  = [\overrightarrow{\mathbf{s}}_{3,k};\overleftarrow{\mathbf{s}}_{3,k}] \in \mathbb{R}^{2n}$).

Now, a weighted averaging mechanism~\cite{bib:bahdanau2014neural} (WAM) as described in Equation~\ref{eq:final_vector} is applied to get a single feature vector $\mathbf{s}\in\mathbb{R}^{2n}$. 
\begin{align}
\label{eq:final_vector}
\mathbf{s} &= \sum_{i=2}^{64}a_i\mathbf{s}_i
\end{align}
\begin{align*}
    \text{where}\quad a_i &= \dfrac{\exp{(\mathbf{s}_i^T\mathbf{w}+b_w)}}{\sum_{j=2}^{64}\exp{(\mathbf{s}_j^T\mathbf{w}+b_w)}}\in(0,1)
\end{align*}

Here, $\mathbf{w}\in\mathbb{R}^{2n}$ and $b_w\in\mathbb{R}$
are trainable parameters of neural network with linear activation and are jointly learned with all the other LSTM parameters during the training process.

Finally the output predicted value $y = \sigma(\mathbf{v}^T\mathbf{s}+b_v)$ is obtained. Here, $\mathbf{v}\in\mathbb{R}^{2n}$ and $b_v\in\mathbb{R}$ are another set of trainable weight parameters, and $\sigma$ is the sigmoid activation function generally used as activation in the last layer for binary classification.

\subsection{Training}
\label{subsec:net_arch}
Proposed model is trained end-to-end to minimize binary cross-entropy loss between the actual labels and the predicted values using Adam optimizer~\cite{bib:kingma2014adam} with the value of $\beta_1$ and $\beta_2$ set at their default values 0.9 and 0.999, respectively. 
Following learning rate strategy is used: an initial learning rate of 0.001 is used for the first ten epochs.
The learning rate is decreased to 0.0005 for the subsequent five epochs and further reduced by a factor of 5 for epoch number sixteen to twenty.
This sequence continues with a 0.00005 learning rate for epochs 21 to 25, and the learning rate is further reduced by a factor of 5 for the subsequent five epochs.
Model is trained for 25 epochs with a batch size of 64.
For increased amount of training data (60\% and 90\% in Section~\ref{subsec:train_size_var}), model is trained for 30 epochs.
The model with the best validation accuracy in these epochs is used for evaluation on the test set.
Our models are trained from scratch without any pre-training or data augmentation. 
For $1\times 1$ convolutions in module $\phi(.)$, kernel weights and biases are initialized using the Glorot uniform initializer~\cite{bib:glorot2010understanding} and zeros, respectively. 
For BiLSTM layers, the kernel weight matrices ($\mathbf{W}$'s) are initialized using the Glorot uniform initializer~\cite{bib:glorot2010understanding}.
Recurrent kernel weights matrices ($\mathbf{V}$'s) are initialized with the orthogonal initialization.
All the bias vectors are initialized with zeroes except for the bias vector corresponding to forget gate, which is initialized with ones.
The value of dropout and recurrent dropout is set to 0.1.

The default value of histogram range ([-b b]), hidden size of each LSTM cell ($n$), and number ($C$) of $1\times1$ filters in the module is experimentally fixed to [-80 80], 128, and 16, respectively. 

\section{Experiments and Discussion}
\label{sec:djpeg_exp}

The first set of experiments reported in this paper are performed on a diverse publicly available dataset~\cite{bib:Park_2018_ECCV}. 
This dataset's diversity lies in the image content and the number of Q-matrices used for generating single and double compressed patches.
Dataset consists of 1,140,430 single and double compressed JPEG patches of size $256\times 256$ generated from 18,946 uncompressed images with 1120 unique Q-matrices (including standard Q-matrices for quality factors 51 - 100). 
These 18,946 uncompressed images were captured by cameras of 15 different models and belonged to three public datasets, RAISE~\cite{bib:dang2015raise}, Dresden~\cite{bib:gloe2010dresden}, and BOSSBase~\cite{bib:bas2011break}.
Single compressed patches were generated by the JPEG compression of each uncompressed patch with a randomly chosen Q-matrix among the 1120 Q-matrices. 
The double compressed patches were generated by re-compressing each of the single compressed patches with another random Q-matrix different from the first Q-matrix.
Further, in the second set of experiments aimed at evaluating the performance on unseen Q-matrices, we also considered separate patches from  four different uncompressed datasets, RAISE~\cite{bib:dang2015raise}, UCID~\cite{bib:schaefer2003ucid}, BOSSBase 1.01~\cite{bib:bas2011break}, and Columbia~\cite{bib:hsu06crfcheck}. 
Similar to the patches in the publicly available dataset~\cite{bib:Park_2018_ECCV}, only the Y channel is quantized while compressing all the images used in this paper. 

Performance evaluation of the proposed system and comparison with the state-of-art systems is done using standard metrics for two-class classification problems. 
Here, double compressed patches are treated as positive class and single compressed patches as negative class. 
Let $tp$ [$tn$] and $fp$ [$fn$] denote the number of patches correctly and incorrectly predicted as double [single] compressed, respectively. 
Then, evaluation metrics used here can be defined as, accuracy = ${(tp+tn)}/{(tp+tn+fp+fn)}$, true positive rate
(TPR) =  ${tp}/{(tp+fn)}$, and true negative rate (TNR) = ${tn}/{(tn+fp)}$.
Hence, TPR [TNR] denotes the fraction of patches correctly detected as double [single] compressed out of the total double [single] compressed patches. 
Since the datasets used for evaluations contain an equal number of single and double compressed patches, metrics such as f-measure are not needed. 

\subsection{Training Size Variation}
\label{subsec:train_size_var}
We analyzed the performance with three different train percentages to address the time complexity in training with a large portion of 1,140,430 patches. 
First, we considered a split where train, validation, and test sets are 30\% (342,130 patches), 10\% (114,044), and 10\%, respectively (30/10/10 split), and the remaining 50\% of the data is not used.
For this 30/10/10 split, we also perform 5-fold cross-validation to check the amount of bias present in a specific randomly chosen 30\% fold.
Table~\ref{tab:train_per_var} shows the results of each of the five-folds. 
Results across these five-folds show minimal variation in performances (with low standard deviation for TPR, TNR, and test accuracy), as evident from the average on the five folds shown in the $7^{th}$ row  (`mean$\pm$ standard deviation'). 
Hence, the computational time is saved by performing all other detailed experiments on a single fold instead of the five folds used here.
For further experiments on the 30/10/10 split, fold-3 data is utilized.
Model hyperparameters are tuned on the validation set of 30/10/10 split and fixed for all other train percentages and other experiments related to the proposed approach. 

We also analyzed the effect of variation in training data size by randomly selecting 30\%, 60\%, and 90\% data as training data. 
The test set of 10\% is kept fixed across all the experiments. 
Corresponding results in Table~\ref{tab:train_per_var} show that when we increase the training data from 30\% to 90\%,  we see a performance improvement of 0.81\%, 0.09\%, and 0.46\% in TPR, TNR, and test accuracy, respectively.
However, this improvement in TPR and test accuracy comes at the cost of around a threefold increase in training time. 
For example, with 30\% training data on NVIDIA GeForce GTX 1080 Ti, the average training time is approximately 1.11 hours per epoch (1.16 days for 25 epochs). 
This training time increases approx two times and three times for 60\% and 90\% train data, respectively. 
Hence, although the proposed system promises to provide further improvement in the performance with 90\% training size, to benefit from shorter training time, and analyze a broader spectrum of various aspects of this forensic problem and the proposed system, except for Section~\ref{subsec:res_analysis} (Table~\ref{tab:90_perf}), all other models are trained on 30\% patches.

\begin{table}[!htb]
	\centering
	\caption{Performance of the proposed system with varying amounts of training data. Bold letter `fold-3' signifies that the train, validation, and test patches corresponding to fold-3 are utilized for futher experiments on 30\% training data scenario.}
	\label{tab:train_per_var} 
	\begin{tabular}{p{0.12\textwidth}p{0.08\textwidth}p{0.08\textwidth}p{0.12\textwidth}}
		\hline\hline
		Train Percentage      & TPR (\%) & TNR (\%) & Test Accuracy (\%)\\ \hline
		30 (fold-1)              & 91.88& 97.57& 94.73 \\ 
		30 (fold-2)               &91.94& 97.93& 94.93 \\ 
		\textbf{30 (fold-3)}               &   91.98 & 97.81 & 94.89 \\ 
		30 (fold-4)               &91.94& 97.77& 94.86 \\ 
		30 (fold-5)               &91.70& 97.95& 94.82 \\ \hline
		30 (5-fold)               &91.89$\pm$0.11&97.81$\pm$0.15&94.85$\pm$0.08 \\ \hline
		60           &92.22& 97.99& 95.10  \\ 
		90         &92.79& 97.90& 95.35\\ \hline\hline
		
	\end{tabular}
\end{table}

\subsection{Comparison with State-of-the-Art Systems}
 \label{subsec:res_analysis}

The proposed system is compared with the following state-of-the-art CNN-based systems: 
Wang \textit{et al.}~\cite{bib:wang2016double},
Barni \textit{et al.}~\cite{bib:barni2017aligned}, 
Verma \textit{et al.}~\cite{bib:verma2018dct}, 
Park \textit{et al.}~\cite{bib:Park_2018_ECCV}, 
Zeng \textit{et al.}~\cite{bib:zeng2019detection}, 
Kwon \textit{et al.}~\cite{bib:kwoncat}.
In~\cite{bib:Park_2018_ECCV}, the performance of various existing methods is reported for 90\% of the training data (1,026,388 patches) and 10\% test data. 
We have adopted the same strategy (randomly selected 90\% patches for training and the remaining 10\% patches for testing).
Results in Table~\ref{tab:90_perf} demonstrate the improvement in overall accuracy and ability to correctly classify single and double compressed patches compared to other approaches.
Since the experimental settings such as datasets is same, Table~\ref{tab:90_perf} shows the results for systems~\cite{bib:wang2016double, bib:barni2017aligned}, and~\cite{bib:Park_2018_ECCV}, as reported in~\cite{bib:Park_2018_ECCV} (also utilized in recent work~\cite{bib:kwoncat}). 
Table~\ref{tab:90_perf} shows the result for system~\cite{bib:kwoncat} as reported in~\cite{bib:kwoncat}. 
The results of the systems~\cite{bib:verma2018dct, bib:zeng2019detection} are generated with our implementation, using the respective papers' default settings. 
In~\cite{bib:zeng2019detection}, the system is trained on the patches of size $64 \times 64$ only, while this comparison needs to be done on the patch size $256\times 256$. 
Thus, our implementation of~\cite{bib:zeng2019detection} uses patches of size $256\times 256$ and 16 as batch size.  

\begin{table}[!htb]
 \centering
         \caption{Performance of various approaches using 90\% patches for training}
    \label{tab:90_perf}
     \resizebox{0.5\textwidth}{!}{
    \begin{tabular}{p{0.2\textwidth}p{0.06\textwidth}p{0.06\textwidth}p{0.12\textwidth}}
        \hline
        System  & TPR (\%) & TNR (\%)& Test Accuracy (\%)\\ \hline
        Wang \textit{et al.}~\cite{bib:wang2016double}  & 67.74 & 78.37& 73.05\\ 
        Barni \textit{et al.}~\cite{bib:barni2017aligned} ([-50 50]) & 77.47 & 89.43 & 83.47 \\ 
        Barni \textit{et al.}~\cite{bib:barni2017aligned} ([-60 60]) & 78.35 & 90.53 & 84.46 \\ 
        Verma \textit{et al.}~\cite{bib:verma2018dct} & 79.15 & 95.68  & 87.41\\ 
        Park \textit{et al.}~\cite{bib:Park_2018_ECCV}  &90.90&94.59 & 92.76\\ 
        		Zeng \textit{et al.}~\cite{bib:zeng2019detection}  &75.80&95.28&85.54\\ 
        		Kwon \textit{et al.}~\cite{bib:kwoncat}  &89.43 &97.75&93.93\\ 
        Proposed System  &\textbf{92.79}&\textbf{97.90}&\textbf{95.35} \\ \hline
    \end{tabular}
}
\end{table}

\subsection{Patch-Size Variation}
\label{subsec:patch_perf}

A system's success in classifying patches of smaller sizes is of great importance in paving the way to utilize that system in localizing forgeries of smaller sizes. 
Table~\ref{tab:patch_perf} shows the performance of different approaches with three different patch sizes. 
The results of~\cite{bib:Park_2018_ECCV} are generated using the available  code\footnote{https://github.com/plok5308/DJPEG} with 100 epochs, 64 batch-size, and learning rate $10^{-5}$.
To ensure the same number of training and test patches as for $256\times 256$ patches, for other patch sizes such as $64\times 64$ and $128 \times 128$, coefficients corresponding to the respective patch sizes are taken from the top left corner of the quantized DCT coefficients of $256\times 256$ patches.
For~\cite{bib:zeng2019detection}, their default batch size of 64 is used while working with $64\times 64$ patches. 
The proposed system consistently outperforms the other three approaches for all the three patch sizes in all three metrics TPR, TNR, and test accuracy.
One important observation that can be made for patch size $256\times 256$, is that for 30\% of training data, improvement in the test accuracy with respect to the baseline system~\cite{bib:Park_2018_ECCV} is more significant (4.59\% in Table~\ref{tab:patch_perf} ) as compared to the case of 90\% training data (2.59\% in Table~\ref{tab:90_perf}).

\begin{table}[!htb]
\centering
\caption{Performance (\%) with patch-size $64\times 64$, $128\times 128$, and $256\times 256$}
\label{tab:patch_perf}
\begin{tabular}{p{0.07\textwidth}p{0.15\textwidth}p{0.04\textwidth}p{0.04\textwidth}p{0.09\textwidth}}
\hline
              Patch-Size & System    & TPR  & TNR & Test Accuracy \\ \hline
\multirow{5}{*}{$64\times 64$}&Verma \textit{et al.}~\cite{bib:verma2018dct} & 66.28 & 91.50 & 78.89\\ 
&Park \textit{et al.}~\cite{bib:Park_2018_ECCV}             & 76.64 & 91.24& 83.94\\ 
&Zeng \textit{et al.}~\cite{bib:zeng2019detection}              &68.19& 89.89& 79.04\\ 
&Proposed System  &\textbf{85.98}& \textbf{93.37}& \textbf{89.68}\\ \hline \hline
\multirow{5}{*}{$128\times 128$}&Verma \textit{et al.}~\cite{bib:verma2018dct}  &77.03  & 88.09 & 82.56\\ 
&Park \textit{et al.}~\cite{bib:Park_2018_ECCV}             & 82.78 & 92.58 &  87.68\\ 
&Zeng \textit{et al.}~\cite{bib:zeng2019detection}             & 75.00& 89.81& 82.41  \\ 
&Proposed System    &\textbf{89.26}& \textbf{96.57}& \textbf{92.92}\\ \hline \hline
\multirow{5}{*}{$256\times 256$}&Verma \textit{et al.}~\cite{bib:verma2018dct}    & 78.52 & 94.89& 86.71
 \\ 
&Park \textit{et al.}~\cite{bib:Park_2018_ECCV}                        & 84.89    & 95.70  &  90.30 \\
&Zeng \textit{et al.}~\cite{bib:zeng2019detection}            & 73.75& 95.78& 84.77  \\ 
&Proposed System    &   \textbf{91.98} & \textbf{97.81} & \textbf{94.89}    \\ \hline
\end{tabular}
\end{table}

\section{Compression Detection for Unseen Q-matrices}
\label{sec:unseen_q}

In most real-life scenarios, the set of potential Q-matrices used for the first compression is unknown beforehand. 
Thus, it is necessary to design systems that can handle unseen Q-matrices as the set of 1,120 Q-matrices is not the complete set of all possible Q-matrices. 
This section aims to analyze the effect of using the same and different (with respect to training) Q-matrices during testing. 
For analyzing Q-matrices' effect only, the same images (in terms of image content) are used for both seen and unseen Q scenarios. 
We also wanted to test the scenario when the patches from a completely different dataset compressed with seen as well as unseen Q-matrices appeared in testing.  
To test this scenario, we consider four different uncompressed datasets, namely RAISE~\cite{bib:dang2015raise}, UCID~\cite{bib:schaefer2003ucid}, BOSSBase 1.01~\cite{bib:bas2011break}, and Columbia~\cite{bib:hsu06crfcheck}. The same procedure as in~\cite{bib:Park_2018_ECCV} is adopted for single and double compressed patch generation from each of these four datasets.
From each uncompressed image, non-overlapping patches of size $256\times256$, starting from top left in the raster order, are extracted, and then single and double compressed test patches are generated. 
RAISE~\cite{bib:dang2015raise} dataset consists of 8156 very high-resolution such as $4928\times3264$, $4288\times2848$, and $3008\times2000$ uncompressed color images captured from three camera models. 
UCID has 1338 original uncompressed color images of resolution $384\times512$ and $512\times384$ captured from the single camera.
Columbia~\cite{bib:hsu06crfcheck} dataset, has 183 authentic uncompressed color images with image size ranging from $757\times568$ to $1152\times768$, captured from four cameras.
BOSSBase 1.01~\cite{bib:bas2011break} has 10,000 never-compressed grayscale images of the size $512\times512$ pixels, taken from seven different cameras.
All images in UCID~\cite{bib:schaefer2003ucid}, BOSSBase 1.01~\cite{bib:bas2011break}, and Columbia~\cite{bib:hsu06crfcheck} datasets are used to create the test patches.
Further, the 1,120 available Q-matrices are divided into two parts.
In the first part, we randomly picked 70\% (784 Q-matrices), while the remaining 30\% (336 Q-matrices) are picked in the second part. 
To test the generalization ability across datasets and unseen Q-matrices, we create the following test datasets: 
1) RAISE Seen Q,
2) RAISE Unseen Q,
3) UCID Seen Q,
4) UCID Unseen Q,
5) BOSSBase 1.01 Seen Q,
6) BOSSBase 1.01 Unseen Q,
7) Columbia seen Q, and 8) Columbia unseen Q.
Here, the word `Seen Q' refers to test patches compressed with the 70\% Q-matrices used in training, while the term `Unseen Q' refers to the same patches compressed with 30\% remaining Q-matrices that are not used in training.

\begin{figure*}[!htb]
\resizebox{\textwidth}{!}{
\begin{tabular}{cccc}
\centering
RAISE~\cite{bib:dang2015raise} & UCID~\cite{bib:schaefer2003ucid} & BOSSBase 1.01~\cite{bib:bas2011break} & Columbia~\cite{bib:hsu06crfcheck}\\
\includegraphics[width=0.28\textwidth,trim={0.26cm 0 1.5cm 0},clip]{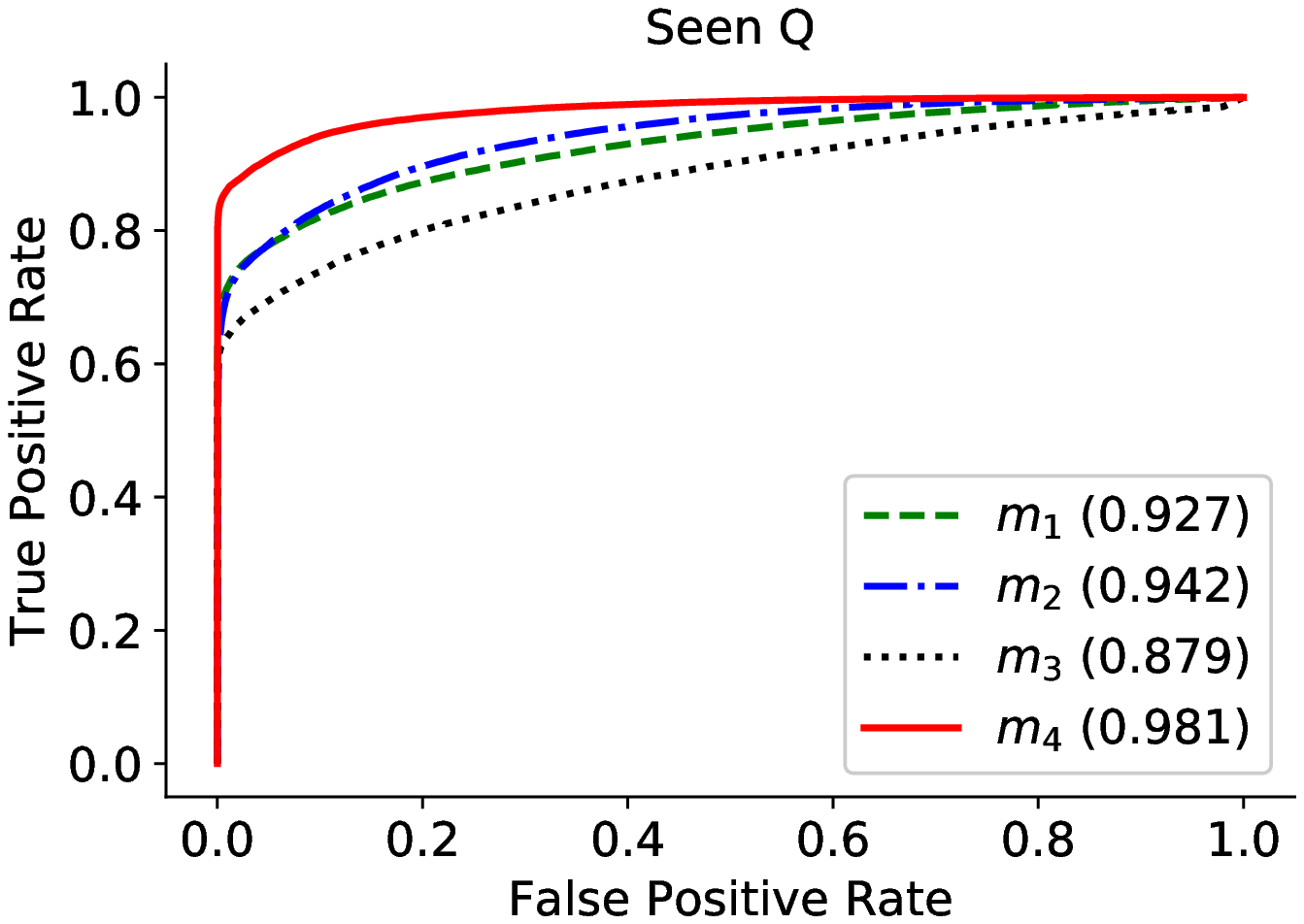} &
\includegraphics[width=0.25\textwidth,trim={1.8cm 0 1.5cm 0},clip]{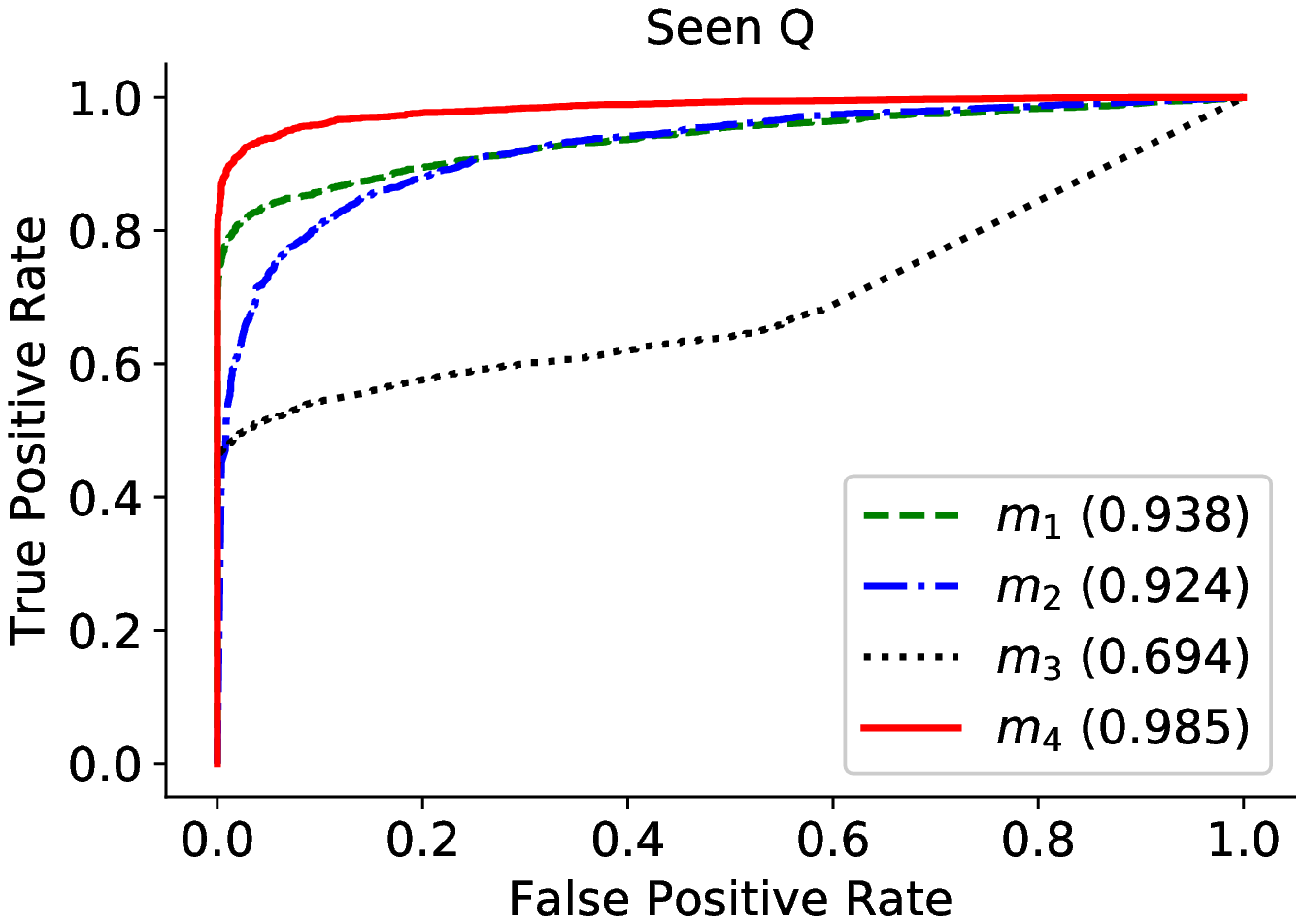}  &
\includegraphics[width=0.25\textwidth,trim={1.8cm 0 1.5cm 0},clip]{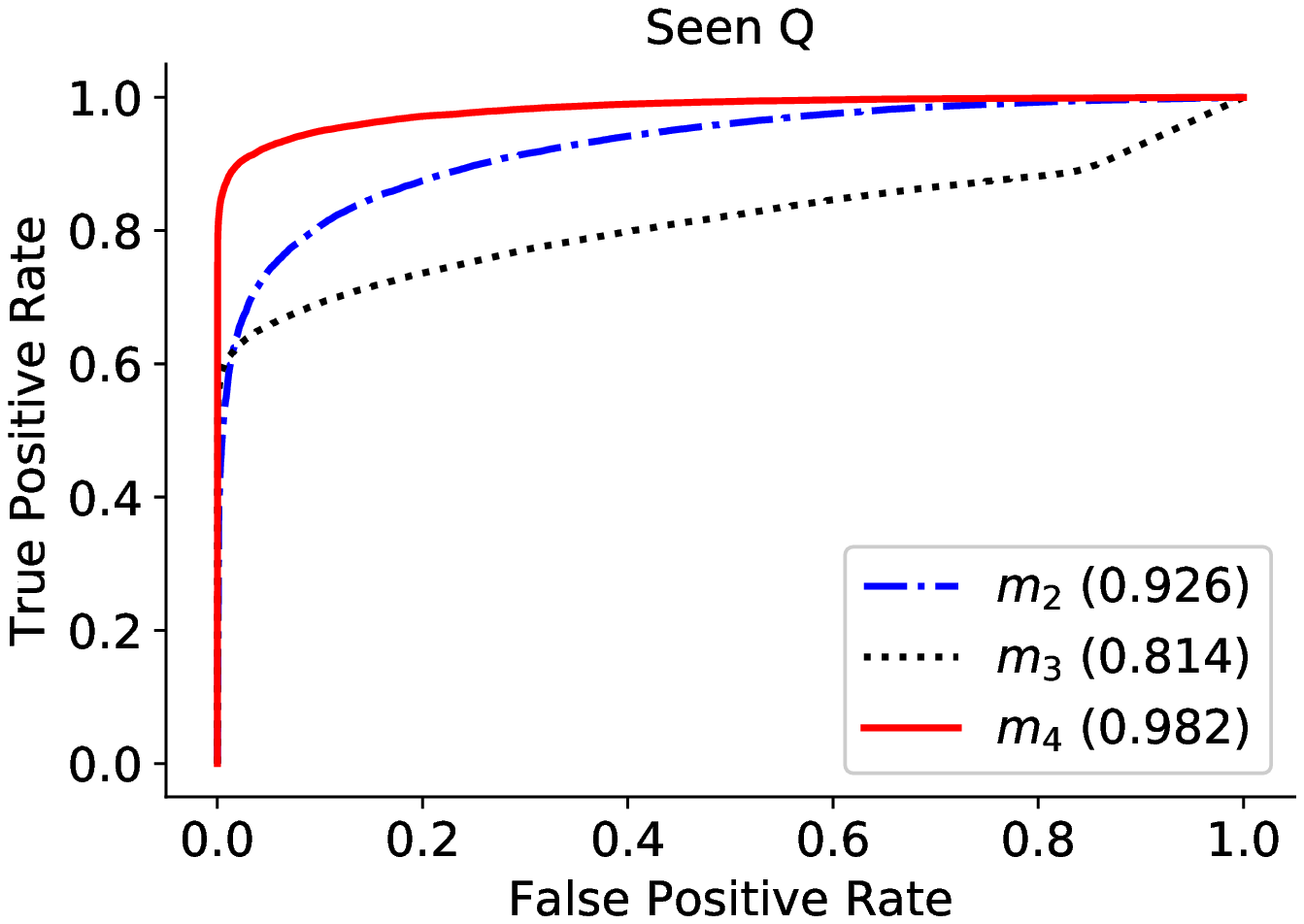}  &
\includegraphics[width=0.25\textwidth,trim={1.8cm 0 1.5cm 0},clip]{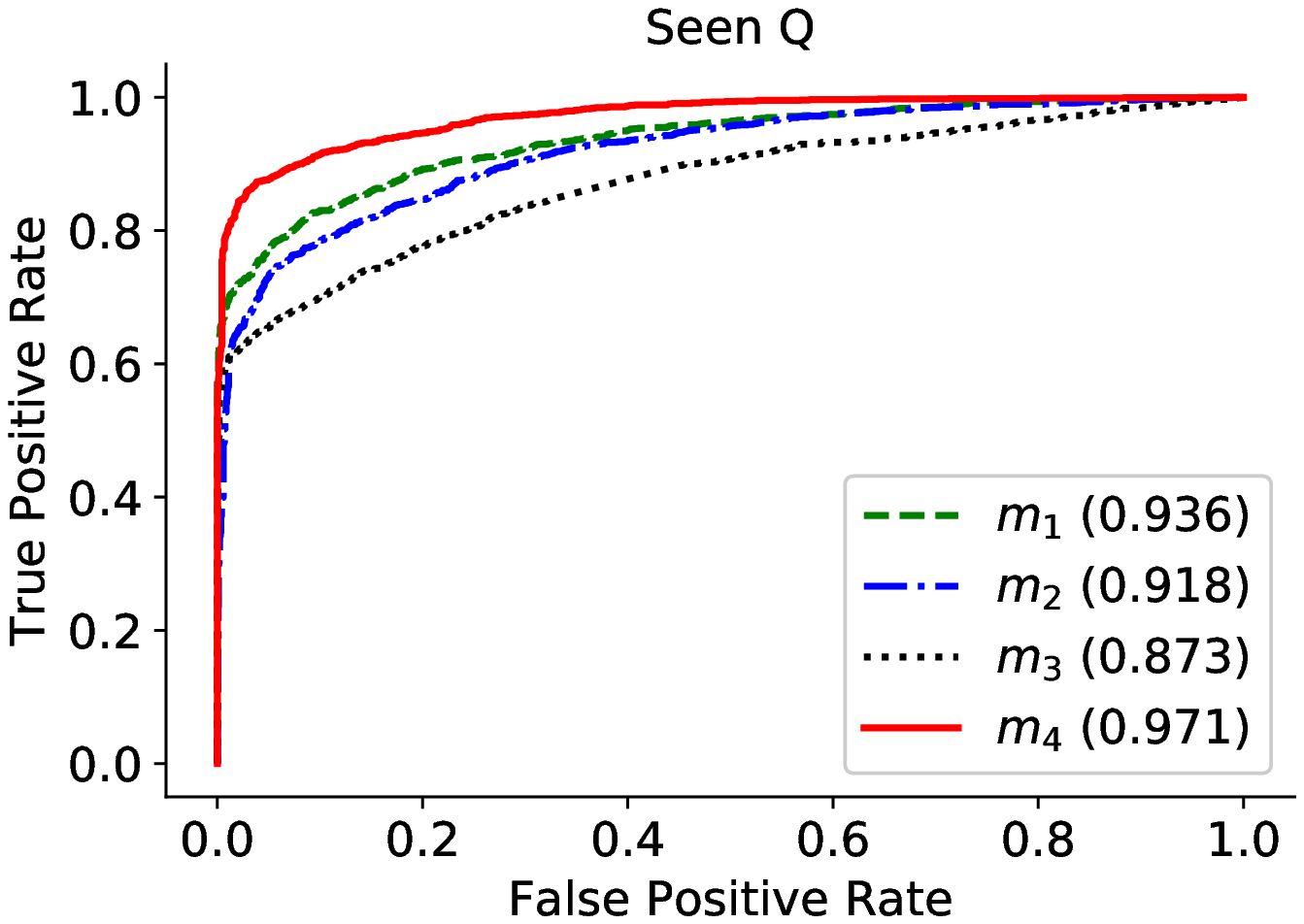} \\

\includegraphics[width=0.28\textwidth,trim={0.26cm 0 1.5cm 0},clip]{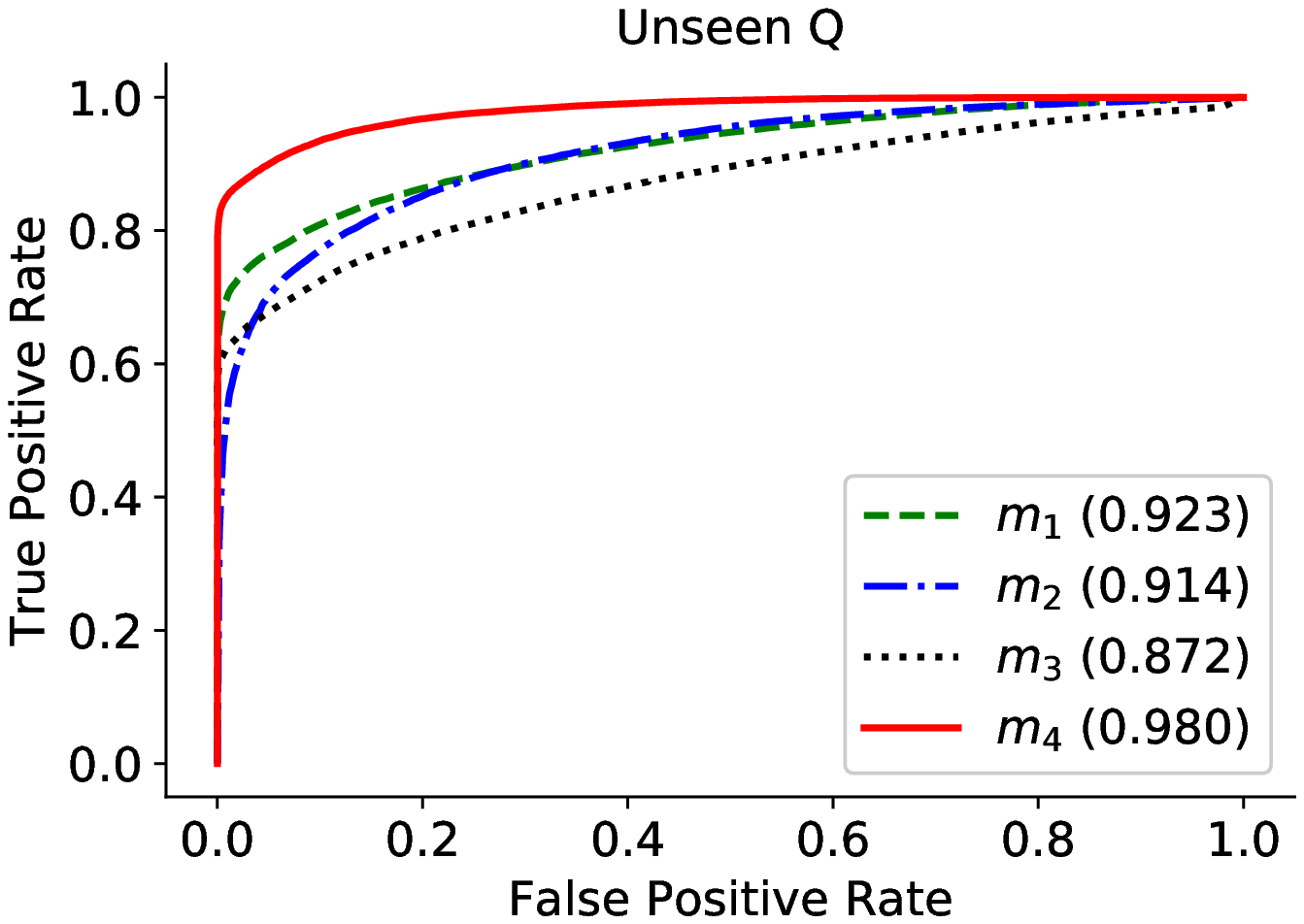} &
\includegraphics[width=0.25\textwidth,trim={1.8cm 0 1.5cm 0},clip]{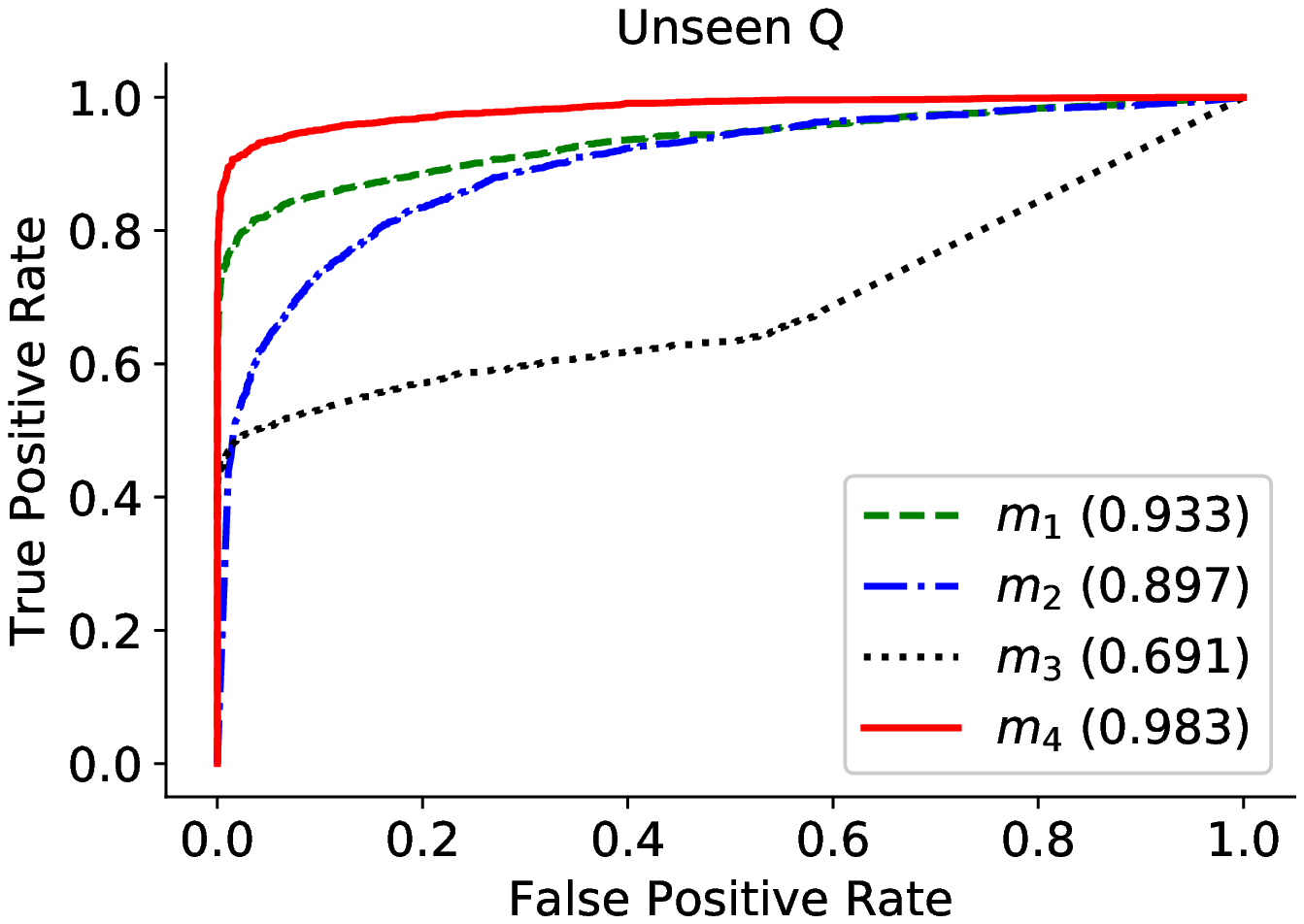}&
\includegraphics[width=0.25\textwidth,trim={1.8cm 0 1.5cm 0},clip]{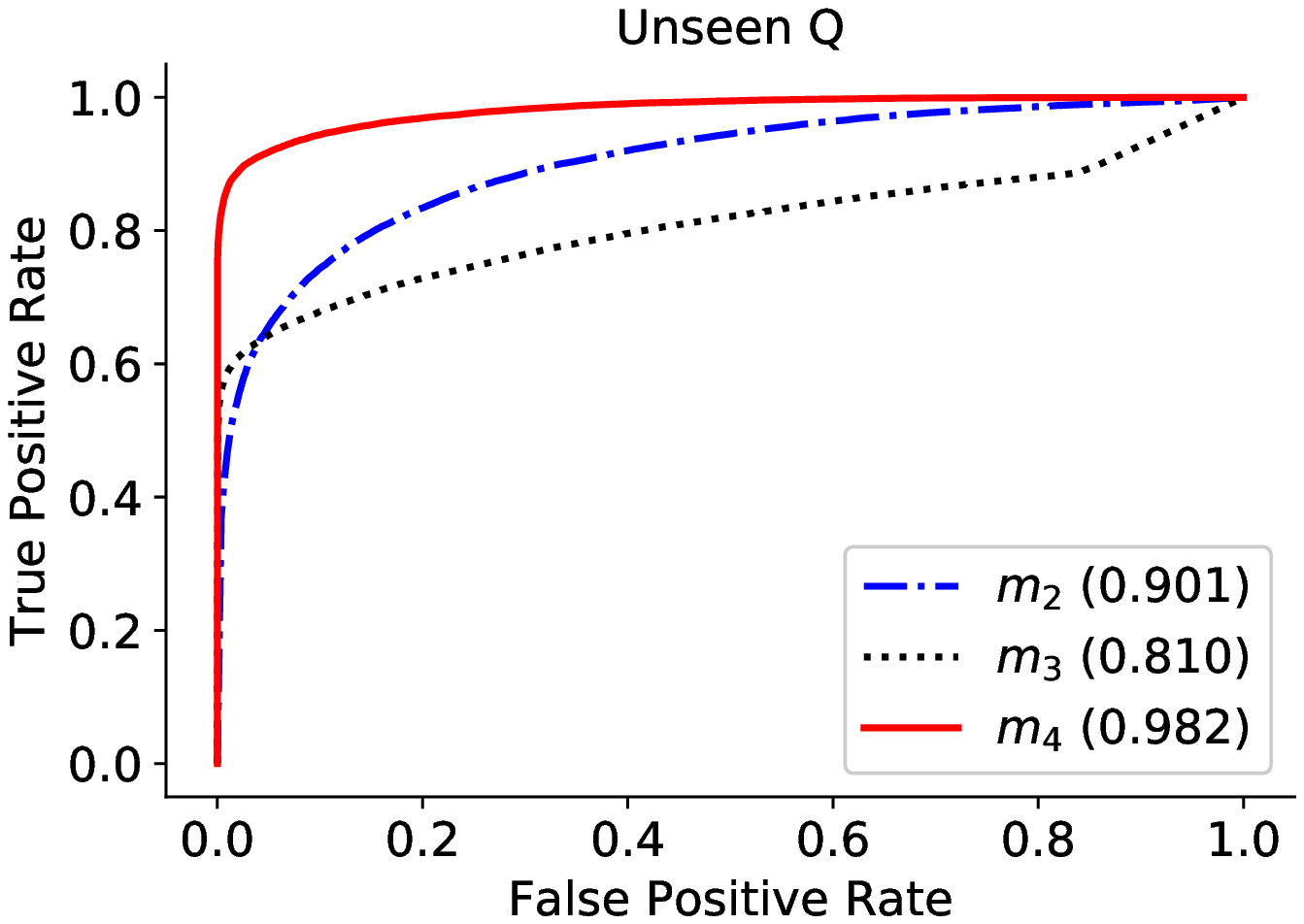}&
\includegraphics[width=0.25\textwidth,trim={1.8cm 0 1.5cm 0},clip]{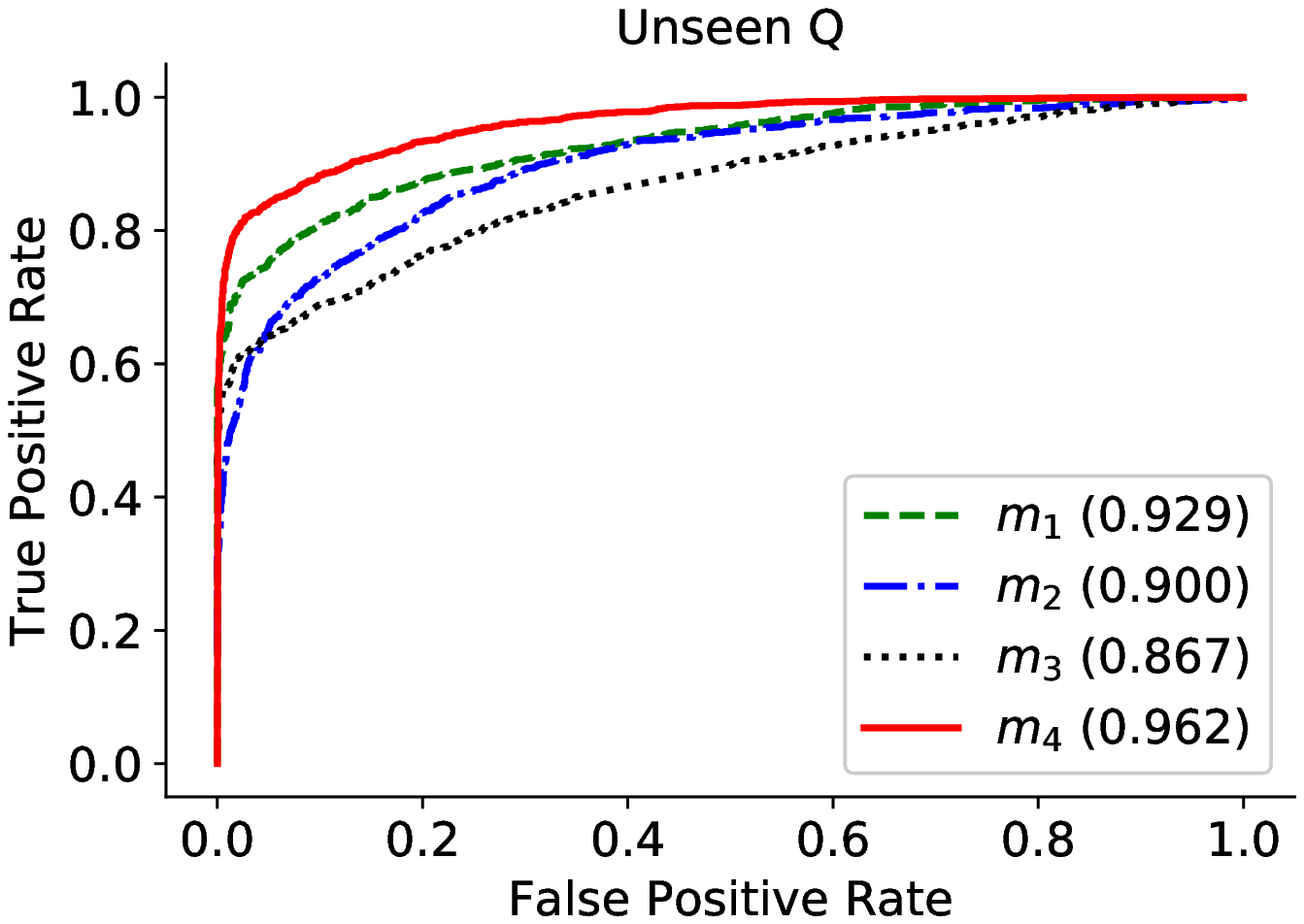} 
\\
&&&\\
\multicolumn{4}{c}{[$m_1$:Verma \textit{et al.}~\cite{bib:verma2018dct}, $m_2$:Park \textit{et al.}~\cite{bib:Park_2018_ECCV}, $m_3$: Zeng \textit{et al.}~\cite{bib:zeng2019detection}, $m_4$: Proposed System]}\\

\end{tabular}
}
\caption{Receiver operating characteristic (ROC) curves. Numbers in the bracket show area under the curve (AUC) for the specific systems.
Model is only trained on the compressed patches generated for the RAISE dataset~\cite{bib:dang2015raise} and evaluated with the RAISE~\cite{bib:dang2015raise} and three other datasets, namely UCID~\cite{bib:schaefer2003ucid}, BOSSBase 1.01~\cite{bib:bas2011break}, and Columbia~\cite{bib:hsu06crfcheck}. The word `Seen Q' refers to test patches compressed with the Q-matrices used in training, while the term `Unseen Q' refers to the same patches compressed with remaining Q-matrices that are not used in training for the respective datasets.}
\label{fig:roc_unseen}
\end{figure*}

A model is trained on single and double compressed patches ($256\times256$) generated from randomly selected images from RAISE~\cite{bib:dang2015raise} dataset using 70\% of the Q-matrices.
The number of training, validation, and test patches for the RAISE~\cite{bib:dang2015raise} dataset is kept the same as the numbers in the original 30/10/10 split used in Section~\ref{subsec:train_size_var}. 
Figure~\ref{fig:roc_unseen} shows the receiver operating characteristic (ROC) curves for all the four datasets in the seen and unseen cases for the proposed system as well as the other three baseline systems. 
Table~\ref{tab:unseen_jpg}, shows the TPR, TNR, and test accuracies for the threshold value of 0.5.
The proposed system consistently performs well across all the scenarios involving unseen Q-matrices and drastically varying image content. 
For example, when the image content of the testing dataset comes from a different distribution as compared to the training dataset (such as UCID~\cite{bib:schaefer2003ucid}, BOSSBase 1.01~\cite{bib:bas2011break}, and Columbia~\cite{bib:hsu06crfcheck}), and the patches are compressed with different Q-matrices as compared to Q-matrices used for training. 
Also, the proposed system significantly outperforms all other baseline systems for seen 
as well as unseen scenarios. 
For example, for the Unseen Q scenario, the proposed system gives 92.50\%, 93.37\%, 92.94\% and 88.60\% accuracies on RAISE, UCID, BOSSBase 1.01, and Columbia datasets, respectively. 
In these four cases, corresponding accuracies for the second best method are 85.66\%, 88.73\%, 81.26\% and 85.54\%.
For the proposed method, comparing the AUC in Figure~\ref{fig:roc_unseen} and the three metrics in Table~\ref{tab:unseen_jpg}, respectively, for 'Seen Q' and 'Unseen Q' scenario, the following key observation can be drawn.
In the proposed method, the drop in performance for the 'Unseen Q' scenario is minimal for all four datasets, indicating the proposed system's robustness to different testing datasets generated from different Q-matrices compared to training.
While for the method in~\cite{bib:Park_2018_ECCV}, a significant drop in performance can be seen in the `Unseen Q' matrices scenario for all the four datasets considered in this section.
This improved overall performance in terms of the TPR, TNR, and test accuracy of the proposed approach suggests the proposed system's applicability as a universal single and double compressed block detector.

\begin{table*}[h!]
  \begin{center}
         \caption{Performance (\%) for the patches compressed from different datasets with seen and unseen quantization matrices. Model is only trained on the compressed patches generated from the RAISE dataset~\cite{bib:dang2015raise} and tested on the RAISE~\cite{bib:dang2015raise} and three other datasets, namely UCID~\cite{bib:schaefer2003ucid}, BOSSBase 1.01~\cite{bib:bas2011break}, and Columbia~\cite{bib:hsu06crfcheck}. The word `Seen Q' refers to test patches compressed with the Q-matrices used in training, while the term `Unseen Q' refers to the same patches compressed with remaining Q-matrices that are not used in training for the respective datasets. (`-': system not applicable for grayscale images.)}
    \label{tab:unseen_jpg}
    \begin{tabular}{p{0.12\textwidth}|p{0.04\textwidth}p{0.04\textwidth}p{0.04\textwidth}|p{0.04\textwidth}p{0.04\textwidth}p{0.04\textwidth}|p{0.04\textwidth}p{0.04\textwidth}p{0.04\textwidth}|p{0.04\textwidth}p{0.04\textwidth}p{0.04\textwidth}}\cline{1-13}
      \multirow{3}{*}{\textbf{System}} & \multicolumn{6}{c|}{RAISE~\cite{bib:dang2015raise}} & \multicolumn{6}{c}{UCID~\cite{bib:schaefer2003ucid}}   \\ \cline{2-13}
      
     & \multicolumn{3}{c|}{Seen Q} & \multicolumn{3}{c|}{Unseen Q}  & \multicolumn{3}{c|}{Seen Q} & \multicolumn{3}{c}{Unseen Q} \\
      
& TPR & TNR & Acc & TPR & TNR & Acc& TPR & TNR & Acc& TPR & TNR & Acc\\ \hline
Verma \textit{et al.}~\cite{bib:verma2018dct}&80.27&92.39&86.33&79.44&91.88&85.66&84.08 &94.54 &89.31 &82.40 & 95.07 &88.73
\\
Park \textit{et al.}~\cite{bib:Park_2018_ECCV}&77.06 &95.72&86.39
&76.62& 90.40& 83.51&87.74& 80.16& 83.95& 88.27& 71.86& 80.06\\
Zeng \textit{et al.}~\cite{bib:zeng2019detection}&70.95&93.52&82.23&69.57&93.03&81.30&51.42 &95.48 &73.45 &50.34& \textbf{95.70}&  73.02\\
Proposed System &\textbf{89.86}&\textbf{95.94}&\textbf{92.90}&\textbf{89.28}&\textbf{95.72}&\textbf{92.50}&\textbf{95.03}& \textbf{92.94}& \textbf{93.98}
&\textbf{94.66}& 92.08& \textbf{93.37}
\\\cline{2-13}
       & \multicolumn{6}{c|}{BOSSBase 1.01~\cite{bib:bas2011break}} & \multicolumn{6}{c}{Columbia~\cite{bib:hsu06crfcheck}}   \\\cline{2-13}
      
Verma \textit{et al.}~\cite{bib:verma2018dct}&-&-&-&-&-&-&82.00&   90.97 &86.49 &79.49& 91.60&  85.54
\\
Park \textit{et al.}~\cite{bib:Park_2018_ECCV}&85.11& 84.39 &84.75& 84.95 &77.58& 81.26&78.69 &89.60  &84.14& 77.77 &85.03 &81.40
\\
Zeng \textit{et al.}~\cite{bib:zeng2019detection}&66.67 &93.89& 80.28& 65.54 &93.53& 79.53&
69.77& 90.23& 80.00&   69.43 &\textbf{88.69}& 79.06\\
Proposed System &\textbf{93.42}& \textbf{93.56}& \textbf{93.49}& \textbf{93.22}& \textbf{92.65}& \textbf{92.94}&\textbf{90.34}& \textbf{91.26}& \textbf{90.80}&  \textbf{89.20}&  88.00  & \textbf{88.60}\\ \hline
    \end{tabular}
  \end{center}
\end{table*}

\section{Ablation Study}
This section analyzes the effect of various network architectural components and design choices on the proposed method's performance.
All the models listed in this section are trained separately from scratch on the patches from the 30/10/10 split's (Section~\ref{subsec:train_size_var}) train set and evaluated on the test set.

We investigate the affect of various histogram bin ranges as shown in Table~\ref{tab:hist_range}.
As the best accuracy is achieved for the histogram bin range $[-80\;80]$, for all other experiments reported in this paper $[-80\;80]$ is used as the histogram bin range. 
Furthermore, it is evident from Table~\ref{tab:hist_range} that even for a bin range of $[-20\;20]$, the test accuracy is close to the best accuracy. 
Therefore, the proposed system can also be used with a smaller bin range, when the system needs to be deployed with reduced computational resources. 
Further, Table~\ref{tab:lstm_layer_nodes} shows the performances with varying number of hidden units, $n\in\{64, 128, 256\}$, used in the forward and backward layers of each LSTM cell at all three layers.
Best performance is observed with the value of $n$ equal to 128.
\begin{table}[!htb]
	\centering
	\caption{Performance comparison with different histogram bin ranges}
	\label{tab:hist_range}
	\begin{tabular}{p{0.12\textwidth}p{0.06\textwidth}p{0.06\textwidth}p{0.12\textwidth}}
		\hline\hline
		Bin range ($[-b\;b]$) & TPR (\%) & TNR (\%) & Test Accuracy (\%)\\ \hline
		$[-20\;20]$         &   91.68  &     97.48& 94.58         \\ 
		$[-40\;40]$         &  91.92   &   97.83  &    94.87      \\ 
		$[-60\;60]$         &  91.91   &   97.56  &  94.73        \\ 
		$[-80\;80]$          &   91.98 & 97.81 & 94.89          \\ 
		$[-100\;100]$            &91.89 & 97.78 & 94.84      \\ \hline
	\end{tabular}
\end{table}
\begin{table}[!htb]
	\centering
	\caption{Performance comparison with different number of hidden units ($n$) in each of the forward and backward LSTM cell}
	\label{tab:lstm_layer_nodes}
	\begin{tabular}{p{0.1\textwidth}p{0.08\textwidth}p{0.08\textwidth}p{0.12\textwidth}}
		\hline\hline
		Units & TPR (\%) & TNR (\%) & Test Accuracy (\%)\\ \hline
		64         &91.52& 97.77& 94.64        \\ 
		128          &   91.98 & 97.81 & 94.89  \\ 
		256            &91.96& 97.70& 94.83      \\ \hline
	\end{tabular}
\end{table}

To recall, in Section~\ref{sec:djpeg_prop_sys}, the proposed architecture contains a non-linear HQ projector module denoted as $\phi(\cdot)$, a stack of two BiLSTM layers (BiLSTM$_1$ and BiLSTM$_2$), a residual connection ($\oplus$), a third BiLSTM layer (BiLSTM$_3$), and a weighted averaging mechanism (WAM).
On the input design side, the proposed system contains histograms and corresponding replicated q-factors for all the 63 AC frequencies.
Table~\ref{tab:ablation1} shows the performance with the various combination of the components except for the fixed histogram range $[-80\;80]$ and the number of hidden units of each LSTM cell set to $128$.
One major observation that can be drawn from the Table~\ref{tab:ablation1}, that even with a single layer BiLSTM (BiLSTM$_1$), with input as a feature sequence obtained from input histograms and corresponding q-factors, method (`model 4') significantly outperforms next best baseline method (test accuracy of 90.30 in Table~\ref{tab:patch_perf} for~\cite{bib:Park_2018_ECCV}) by 3.79\% in test accuracy.
Adding more BiLSTM layers such as BiLSTM$_2$ and BiLSTM$_3$ with a residual connection at each frequency index improves performance.
For the proposed method (`model 10'), non-linear HQ projector module $\phi(\cdot)$ has $C=16$ filters resulting in the best test accuracy of 94.89\%.
The effect of varying the value of $C$ (`model 9' to `model 13') suggests that with a reasonable value of $C$, the variant models' performance is robust to its value. 
Furthermore, we observe that deeper architecture such as adding one more BiLSTM layer such as adding BiLSTM$_4$ without residual connection (`model 14' and `model 15') and residual connection (`model 16' and `model 17') does not further help in performance improvement.

\subsection{Evaluation with existing CNN architectures}
\label{subsec:perf_cnn}
In literature, histograms or certain handcrafted features from the histogram of quantized DCT coefficients or de-quantized DCT coefficients are used as input to CNN to detect the re-compression.
Recent methods such as~\cite{bib:wang2016double, bib:verma2018dct} use 1D concatenated histogram from selected low-frequency locations (in zig-zag order~\cite{bib:wallace1992jpeg}). 
While other methods such as~\cite{bib:barni2017aligned,bib:Park_2018_ECCV} use frequency-wise stacked histograms or features from histogram as input to the respective 2D convolutional neural networks.
Exact comparative performances with these methods~\cite{bib:wang2016double, bib:verma2018dct,bib:barni2017aligned,bib:Park_2018_ECCV} are reported in Table~\ref{tab:90_perf}. 
To this en route, in this section, we investigated the performances of our raw histograms' and proposed way of combining it with the respective q-factors with the common CNN architectures.
Architectures considered in this study are ResNet-50~\cite{bib:he2016deep}, ResNet-152~\cite{bib:he2016deep}, DenseNet-121~\cite{bib:huang2017densely} with growth-rate 32 and 40.
For better comparative analysis, performances with these architectures are reported in the lower half of Table~\ref{tab:ablation1} (`model 18' to `model 25').
To accomplish this task, for the histogram inputs, we vertically stack the histograms of 63 frequencies in raster order of the frequency index from that is left to right and top to bottom, which results in input dimension of $63\times (2b+1)$ that is fed to the respective CNN architectures.
While in the case of the proposed way of using q-factors alongside the histogram vectors of each 63 frequencies, we use a channel-wise concatenation of histogram vectors and respective vectors obtained from repeating the q-factors, that results in input dimension of ${63\times (2b+1)\times 2}$. This input combination results in better test accuracy as compared to histogram-only input.
Comparing these CNN models with the proposed BiLSTM based model, even the simplest BiLSTM based model (`model 4') with $0.3$ million parameters performs better in all three metrics than best performing DenseNet-121 (`model 25') with $10.87$ million parameters.

\begin{table*}[!htb]
	\centering
	\caption{Ablation studies with various hyper-parameters of the proposed model and other CNN architectures. All the models listed here are trained separately from scratch. Performances are reported on the test set. $\oplus$ denotes the residual connection between BiLSTM$_2$ and BiLSTM$_3$. WAM: weighted averging mechanism. $^*$: $\phi(\cdot)$ contains the following sequence of operations; one $1\times1$ convolution, BN, and ReLU.}. 
	\label{tab:ablation1}
	\begin{tabular}{p{0.06\textwidth}|p{0.02\textwidth}|p{0.05\textwidth}|p{0.04\textwidth}|p{0.06\textwidth}p{0.06\textwidth}p{0.02\textwidth}p{0.06\textwidth}p{0.06\textwidth}p{0.03\textwidth}|p{0.05\textwidth}p{0.05\textwidth}p{0.05\textwidth}p{0.06\textwidth}}
		\hline
	 \multirow{ 2}{*}{Model}&\multicolumn{2}{c|}{Input Design} &$\phi(\cdot)$& \multicolumn{6}{c|}{Feature Encoder} & \multirow{ 2}{*}{TPR (\%)} & \multirow{ 2}{*}{TNR (\%)} &  \multirow{ 2}{*}{Acc (\%)} &\multirow{ 2}{*}{\#Params}\\ \cline{2-3} \cline{5-10} 
		& hist&q-factor& C=16&BiLSTM$_1$&BiLSTM$_2$&$\oplus$& BiLSTM$_3$&BiLSTM$_4$& WAM & &&&\\
		\hline
model 1 &\cmark&&  &  &  &  &  & & &59.51& 81.04&70.28&10,144\\ 
model 2 &\cmark&\cmark & \cmark &  &  &  &  & & &69.06&89.35&79.21&10,283\\\hline
model 3 &\cmark& &  & \cmark &  &  &  & & \cmark &81.12& 96.40& 88.76& 297,474\\ 
model 4 &\cmark&\cmark & \cmark & \cmark &  &  &  & & \cmark &90.64& 97.54& 94.09& 297,613\\ \hline
model 5 &\cmark& &  & \cmark & \cmark &  &  & &\cmark &84.77& 97.20& 90.99& 691,714\\ 
model 6 &\cmark&\cmark & \cmark & \cmark & \cmark &  &  & &\cmark &91.68& 97.83& 94.76& 691,853\\ \hline
model 7 &\cmark& &  & \cmark & \cmark &  & \cmark & &\cmark &85.87& 97.16& 91.51& 1,085,954\\ 
model 8 &\cmark&\cmark & \cmark & \cmark & \cmark &  & \cmark & &\cmark &91.40& 97.98& 94.69& 1,086,093\\ \hline
model 9 &\cmark& &  & \cmark & \cmark & \cmark & \cmark & &\cmark &85.40& 97.02& 91.21& 1,085,954\\ 
\textbf{model 10} &\cmark&\cmark & \cmark & \cmark & \cmark & \cmark & \cmark & &\cmark &91.98& 97.81& \textbf{94.89}& 1,086,093\\ 
model 11 &\cmark&\cmark & C=1$^*$ & \cmark & \cmark & \cmark & \cmark & &\cmark &91.53& 97.73& 94.63& 1,085,961\\ 
model 12 &\cmark&\cmark & C=8 & \cmark & \cmark & \cmark & \cmark & &\cmark &92.11& 97.35& 94.73& 1,086,029\\ 
model 13 &\cmark&\cmark & C=32 & \cmark & \cmark & \cmark & \cmark & &\cmark &91.69& 97.93& 94.81& 1,086,221\\  \hline
model 14 &\cmark& &  & \cmark & \cmark &  & \cmark & \cmark&\cmark &86.01& 97.51& 91.76& 1,480,194\\ 
model 15 &\cmark&\cmark & \cmark & \cmark & \cmark &  & \cmark & \cmark&\cmark &91.10& 98.25& 94.67& 1,480,333\\ \hline
model 16 &\cmark& &  & \cmark & \cmark & \cmark & \cmark & \cmark&\cmark &86.12& 96.93& 91.53
& 1,480,194\\ 
model 17 &\cmark&\cmark & \cmark & \cmark & \cmark & \cmark & \cmark & \cmark & \cmark &91.64 & 98.00& 94.82& 1,480,333\\ \hline\hline
& & & \multicolumn{7}{c|}{CNN Architectures with Histograms and Proposed Input} &  &  & & \\ \hline\hline
model 18&\cmark&& \multicolumn{7}{c|}{ResNet-50~\cite{bib:he2016deep}} &82.77& 98.51& 90.64& 23,583,489\\
model 19&\cmark&\cmark& \multicolumn{7}{c|}{ResNet-50~\cite{bib:he2016deep}} &89.72& 97.33& 93.52& 23,586,625\\\hline 
model 20&\cmark&& \multicolumn{7}{c|}{ResNet-152~\cite{bib:he2016deep}} &85.34& 96.22& 90.78& 58,366,721\\
model 21&\cmark&\cmark& \multicolumn{7}{c|}{ResNet-152~\cite{bib:he2016deep}} &90.81& 96.31& 93.56& 58,369,857\\\hline 
model 22&\cmark&& \multicolumn{7}{c|}{DenseNet-121~\cite{bib:huang2017densely} with growth-rate 32} &83.76& 97.08& 90.42& 7,038,145\\
model 23&\cmark&\cmark& \multicolumn{7}{c|}{DenseNet-121~\cite{bib:huang2017densely} with growth-rate 32}&90.02& 97.20& 93.61& 7,041,281 \\\hline 
model 24&\cmark&& \multicolumn{7}{c|}{DenseNet-121~\cite{bib:huang2017densely} with growth-rate 40}&81.82& 98.69& 90.25& 10,873,081 \\
model 25&\cmark&\cmark& \multicolumn{7}{c|}{DenseNet-121~\cite{bib:huang2017densely} with growth-rate 40} &89.87& 97.42& 93.65& 10,876,217 \\\hline

	\end{tabular}
\end{table*}

\subsection{Variations in feature encoder}
\subsubsection{Encoding with zig-zag order of frequency} 
Some of the JPEG compression detection algorithms~\cite{bib:wang2016double,bib:amerini2017localization,bib:zeng2019detection} select DCT coefficients from some low-frequency locations in zig-zag~\cite{bib:wallace1992jpeg} order to get sufficient statistics; as the higher frequency locations are quantized heavily during JPEG compression. 
To investigate if selecting the coefficients in zig-zag order affects the proposed system's compression detection performance, we trained a model where histograms and corresponding q-factors are passed in zig-zag order, keeping all other settings the same as the initially proposed model. 
We observed that the model's performance in terms of (TPR, TNR, Accuracy) was on the lower side (91.50, 97.14, 94.32) compared to the proposed model (91.98, 97.81, 94.89).

\subsubsection{Without weighted averaging mechanism}
In Table~\ref{tab:wo_wt_mech}, we show the performances without using the weighted averaging mechanism (WAM) in the proposed approach.
In this case, we will have the sequence vectors obtained from the BiLSTM$_3$.
One possible approach can be to use the forward and backward concatenated hidden state vectors at frequency index $64$, that is $\mathbf{s}_{64}$ ($\mathbf{s}_{64} = [\overrightarrow{\mathbf{s}}_{3,64};\overleftarrow{\mathbf{s}}_{3,64}]$) as a summarized feature vector.
Alternatively, similar idea can be applied at the frequency index $2$ with $\mathbf{s}_{2}= [\overrightarrow{\mathbf{s}}_{3,64};\overleftarrow{\mathbf{s}}_{3,64}]$, due to the utilization of BiLSTM cells in the proposed network.
Further, these two vectors at index $2$ and $64$ ($\mathbf{s}_{2}$ and $\mathbf{s}_{64}$) can be element-wise added ($\mathbf{s}_{2} \oplus \mathbf{s}_{64} \in \mathbb{R}^{2n}$) or concatenated ($[\mathbf{s}_{2} ; \mathbf{s}_{64}] \in \mathbb{R}^{4n}$), before the final classification.
The performances with these combinations are on the slightly lower side in all three metrics than the proposed method; still, these numbers suggest that the BiLSTM based histogram feature encoding is robust to design choices and is a powerful tool for single vs. double compression detection.

\begin{table}[!htb]
 \centering
         \caption{Performance (\%) without weighted averaging mechanism. }
    \label{tab:wo_wt_mech}
    \begin{tabular}{p{0.2\textwidth}p{0.04\textwidth}p{0.04\textwidth}p{0.09\textwidth}}
        \hline
          Input to Classification Block & TPR & TNR & Test Accuracy\\ \hline
         $\mathbf{s}_{64} = [\overrightarrow{\mathbf{s}}_{3,64};\overleftarrow{\mathbf{s}}_{3,64}] \in \mathbb{R}^{2n}$ &91.78& 97.74& 94.76\\ 
         $\mathbf{s}_{2} = [\overrightarrow{\mathbf{s}}_{3,2};\overleftarrow{\mathbf{s}}_{3,2}] \in \mathbb{R}^{2n}$ &91.72& 97.72& 94.72\\ 
        $\mathbf{s}_{2} \oplus \mathbf{s}_{64} \in \mathbb{R}^{2n}$ &91.72& 97.90& 94.81\\ 
        $[\mathbf{s}_{2} ; \mathbf{s}_{64}] \in \mathbb{R}^{4n}$ &92.02& 97.62& 94.82\\  \hline
    \end{tabular}

\end{table}

\section{Conclusions}
\label{sec:djpeg_conc}

In this paper, we have proposed a deep BiLSTM based framework that uses the features from the histogram of quantized DCT coefficients directly extracted from JPEG file with the corresponding q-factors to effectively capture the compression specific artifacts for classifying JPEG patches as single or double compressed.
With the analysis of the various aspects of the proposed system, we can conclude that the sequence-based modeling of the features obtained from the combination of histograms and respective q-factors provides robust BiLSTM based design choices.
This work is the first attempt to model the histogram-based feature sequences of AC frequencies as sequential data and selectively encode the useful information from the different indices using the sequence models, BiLSTM, in the present case.
We showed significant improvement over the current state-of-the-art methods to classify JPEG blocks as single or double compressed in challenging but very important and realistic scenarios of unseen compression.
In the future direction, one can explore such sequence models with more sophisticated modeling. 

\section*{Acknowledgments}
This material is based upon work partially supported by a grant from the Department of Science and Technology (DST), New Delhi, India, under Award Number ECR/2015/000583. 
Any opinions, findings, and conclusions or recommendations expressed in this material are those of the author(s) and do not necessarily reflect the views of the funding agencies. 

\bibliographystyle{IEEEtran}
\bibliography{djpeg_forgery.bib}

\end{document}